\documentclass[11pt,a4paper]{article}
\usepackage{jheppub}
\pdfoutput=1

\usepackage{siunitx}
\usepackage{pdfpages}
\usepackage{float}
\usepackage{textcomp}               
\usepackage{braket,slashed,bm}
\usepackage{array,multirow, makecell}
\usepackage[normalem]{ulem}
\usepackage[usenames,dvipsnames]{xcolor}
\usepackage{cancel,youngtab}

\usepackage{sectsty}
\allsectionsfont{\boldmath}

\usepackage[T1]{fontenc} 
\usepackage{mathrsfs}
\usepackage{booktabs}
\usepackage{adjustbox}
\usepackage{mathtools}
\usepackage{hhline}
\usepackage{soul}
\usepackage{subcaption}
\usepackage[section]{placeins}

\usepackage{tikz}
\usetikzlibrary{arrows,decorations.pathmorphing,backgrounds,positioning,fit,petri,automata,shadows,calendar,mindmap,
decorations.markings,calc}

\definecolor{labelkey}{rgb}{0,0.5,0.0}

\usepackage{xspace}
\usepackage{slashed}
\usepackage{array,multirow}
\usepackage{graphicx}
\usepackage{graphics}
\usepackage{epsfig}
\usepackage{amsfonts}
\usepackage{amsmath}
\usepackage{amssymb}
\usepackage{dsfont}
\usepackage{color}
\usepackage{pdflscape}
\usepackage{pifont}
\usepackage{pgfplots}
\usepackage{tikz} 
\usepackage{tikz-feynhand}
\usepackage{pgfplotstable}
\usepackage{bbold}
\usepackage[capitalise, english]{cleveref}
\usepackage{diagbox}

\graphicspath{{figs/}}

\newcommand{\lam}{\lambda}

\usepackage{bm}                                  

\newcommand{\beq}{\begin{equation}}
\newcommand{\eeq}{\end{equation}}
\newcommand{\be}{\begin{equation}}
\newcommand{\ee}{\end{equation}}
\newcommand{\bea}{\begin{eqnarray}}
\newcommand{\eea}{\end{eqnarray}}
\newcommand{\ben}{\begin{eqnarray*}}
\newcommand{\een}{\end{eqnarray*}}

\newcommand{\bma}{\begin{pmatrix}}
\newcommand{\ema}{\end{pmatrix}}

\renewcommand{\Re}[1]{{{\rm Re\,} #1}}

\def\lixo#1{}

\def\slashchar#1{\setbox0=\hbox{$#1$}           
  \dimen0=\wd0                                    
  \setbox1=\hbox{/} \dimen1=\wd1                  
  \ifdim\dimen0>\dimen1                           
    \rlap{\hbox to \dimen0{\hfil/\hfil}}            
    #1                                             
  \else                                          
    \rlap{\hbox to \dimen1{\hfil$#1$\hfil}}        
    /                                           
 \fi}                                           %

\newsavebox\foobox 
\newlength{\foodim}
\newcommand{\slantbox}[2][0]{\mbox{%
		\sbox{\foobox}{#2}%
		\foodim=#1\wd\foobox
		\hskip \wd\foobox
		\hskip -0.5\foodim
		\pdfsave
		\pdfsetmatrix{1 0 #1 1}%
		\llap{\usebox{\foobox}}%
		\pdfrestore
		\hskip 0.5\foodim
}}
\def\lgr{\slantbox[-.45]{$\mathscr{L}$}}

\newcommand{\mchi}{m_{\tilde\chi_1^0}}

\renewcommand{\d}{\text{d}}

\newcommand\gsim{\gtrsim}

\newcommand{\dslash}[1]{#1 \llap{/\kern-0.5pt}}
\newcommand{\Dslash}[1]{#1 \llap{/\kern+1.5pt}}
\newcommand{\DDslash}[1]{#1 \llap{/\kern+2.3pt}}
\newcommand{\dslashh}[1]{#1 \llap{/\kern+1pt}}

\definecolor{ForestGreen}{rgb}{0.0, 0.27, 0.13}
\definecolor{newgreen}{HTML}{009900}

	\preprint{\begin{flushright} BONN-TH-2023-06
	\end{flushright}}	
	
	\title{Recasting Bounds on Long-lived Heavy Neutral Leptons in Terms of a Light
           Supersymmetric R-parity Violating Neutralino}

		\author[a]{Herbi~K.~Dreiner,}
		\emailAdd{dreiner@uni-bonn.de}
		\affiliation[a]{Bethe Center for Theoretical Physics \& Physikalisches Institut der Universit\"at Bonn,\\ Nu{\ss}allee 12, 53115 Bonn, Germany}

		\author[a]{Dominik~K\"ohler,}
		\emailAdd{koehler@physik.uni-bonn.de}

		\author[a]{Saurabh~Nangia,}
		\emailAdd{nangia@physik.uni-bonn.de}

            \author[a]{Martin~Sch\"urmann,}
            \emailAdd{marschu@uni-bonn.de}

		\author[b,c]{and Zeren~Simon~Wang}
		\emailAdd{wzs@mx.nthu.edu.tw}
		\affiliation[b]{Department of Physics, National Tsing Hua University, Hsinchu 300, Taiwan}
		\affiliation[c]{Center for Theory and Computation, National Tsing Hua University, Hsinchu 300, Taiwan}

\abstract{In R-parity-violating (RPV) supersymmetric models, light neutralinos with masses from the GeV-scale down to even zero are still allowed by all laboratory constraints. They are further consistent with dark matter observations, as they decay via RPV couplings. These RPV couplings 
are in general constrained to be small. Hence, such light neutralinos, if produced, 
\textit{e.g.}, at a beam-dump or collider experiment, appear as displaced vertices or missing 
energy at the detector level. The same signatures have been extensively searched for at various 
experiments in the theoretical context of sterile neutrinos which mix with active neutrinos.
In this work, we recast the sensitivity of both past and present experiments to sterile neutrinos 
to obtain new bounds on RPV couplings associated with a light neutralino. We find experiments such 
as \texttt{T2K}, \texttt{BEBC}, \texttt{FASER}, \texttt{DUNE}, and \texttt{MoEDAL-MAPP} can improve the current bounds on RPV couplings by up to $3-4$ orders of magnitude in several 
benchmark scenarios.
}

\pgfplotsset{compat=1.8}
\begin{document}
\maketitle

\section{Introduction}\label{sec:intro}

With the discovery of a Standard-Model (SM)-like Higgs boson at the LHC in 2012~\cite{ATLAS:2012yve,CMS:2012qbp}, the 
SM is complete. Yet many questions beyond the Standard Model (BSM) remain. One avenue of exploration which has
received considerable attention is new light, feebly interacting particles~\cite{Dedes:2001zia, Feng:2014uja, 
Hewett:2014qja, Alexander:2016aln, Battaglieri:2017aum, Alimena:2019zri, Beacham:2019nyx, Arguelles:2019xgp, 
EuropeanStrategyforParticlePhysicsPreparatoryGroup:2019qin, Agrawal:2021dbo}. Such exotic states are predicted in many
BSM theories and are often long-lived. Theoretical candidates for such long-lived particles (LLPs) range from heavy 
neutral leptons (HNLs), axion-like particles, dark scalars, and dark photons, to electroweakinos in variations of 
supersymmetric models, inelastic dark matter (DM), and many more. See, \textit{e.g.},~Refs.~\cite{Alimena:2019zri, 
Lee:2018pag, Curtin:2018mvb, Knapen:2022afb} for reviews on LLPs. They are usually motivated as explanations of either
the non-vanishing active neutrino masses or of dark matter and could have a spin of $0,\frac{1}{2},1,
\ldots$, and a mass usually ranging from the sub-MeV scale up to the multi-TeV scale. 

As an example, the HNLs (labeled as $N$ in this work) are proposed hypothetical spin-half fermions 
that are SM-singlets which mix with the light active neutrinos. For certain mass values, they can explain 
simultaneously the neutrino masses, the observed dark matter, as well as the baryon asymmetry of the 
Universe~\cite{Asaka:2005pn}. They can give light neutrinos Dirac masses via the Yukawa term $LHN$ in the Lagrangian with $L$ and $H$ being the lepton and Higgs doublets.
This implies unnaturally small Yukawa couplings given the tiny neutrino masses~\cite{deSalas:2020pgw, 
Esteban:2020cvm, Capozzi:2021fjo}.
One can also write down a Majorana mass term in the Lagrangian, leading to light Majorana neutrinos via the seesaw mechanisms~\cite{Minkowski:1977sc, Yanagida:1979as, Gell-Mann:1979vob, Mohapatra:1979ia, Schechter:1980gr}.
While the vanilla type-I seesaw mechanism demands the mixing parameters should be small for the tiny active neutrino masses, larger values of mixing are legitimately conceived in other variations such as the linear seesaw model~\cite{Akhmedov:1995ip, Akhmedov:1995vm, Malinsky:2005bi} and inverse seesaw model~\cite{Mohapatra:1986bd, Gonzalez-Garcia:1988okv}.
Therefore, in phenomenological studies, the HNL mass and the mixing angles with the SM neutrinos are often assumed to be independent parameters. 

Via mixing with the active neutrinos, the HNLs can participate in both charged-current and 
neutral-current interactions, coupled to both gauge bosons directly and the Higgs boson indirectly. 
They can thus be produced from decays of these bosons or from mesons, or through direct production 
at colliders. The HNL can decay leptonically or semi-leptonically, leading to a variety of signatures 
at the different experimental facilities. In particular, GeV-scale HNLs have received substantial 
attention in recent years, for they could originate from rare decays of mesons, which are copiously 
produced, \textit{e.g.}, at beam-dump experiments, $B$-factories, and high-energy hadron colliders. Given the 
strict experimental upper bounds on the mixing of the HNLs and the active neutrinos, the more recent 
focus has been on small mixing angles, for which the GeV-scale HNLs are usually long-lived. Searches 
for these long-lived HNLs have been performed via many different signatures, including searches for 
missing energy, peak searches, as well as searches for displaced vertices (DV). See, 
\textit{e.g.},~Refs.~\cite{hnl_limits, Fernandez-Martinez:2023phj} for summaries of these searches. 
Moreover, one could use the uncertainty on the measurements of the invisible decay width of 
mesons to put upper bounds on the long-lived HNLs, which contribute to the invisible decay width.

Besides the HNLs, supersymmetric electroweakinos, including charginos and neutralinos, are often considered as LLP candidates. See for instance Refs.~\cite{ATLAS:2022rme, CMS:2020atg}.
In particular, a specific type of light neutralino in the GeV mass scale is still allowed by all observational and experimental 
constraints~\cite{Choudhury:1999tn, Gogoladze:2002xp, Dreiner:2009er, Dreiner:2009ic, Grifols:1988fw, Ellis:1988aa, 
Lau:1993vf, Dreiner:2003wh, Dreiner:2006gu, Profumo:2008yg, Dreiner:2011fp, Dreiner:2013tja}; they are necessarily 
bino-like~\cite{Dreiner:2009ic,Choudhury:1999tn} and have to decay to avoid overclosing the Universe~\cite{Hooper:2002nq,
Calibbi:2013poa, Bechtle:2015nua}. One possibility is to consider R-parity-violating supersymmetry (RPV-SUSY) (see 
Refs.~\cite{Dreiner:1997uz,Barbier:2004ez,Mohapatra:2015fua} for reviews), where the light binos decay via small but 
non-vanishing RPV couplings (see Ref.~\cite{Domingo:2022emr} for a detailed study of light bino decays).
The minimal version is known as the R-parity-violating Minimal Supersymmetric Standard Model (RPV-MSSM) \cite{Allanach:2003eb}. The 
RPV-MSSM solves the SM hierarchy problem as in the MSSM, and also predicts a very rich phenomenology at 
colliders~\cite{Dreiner:1991pe, deCampos:2007bn, Dercks:2017lfq, Dreiner:2023bvs}.
\textit{A priori} it is unknown if R-parity is conserved or broken, SUSY models with R-parity conservation or violation are equally legitimate, \textit{e.g.}, see 
Refs.~\cite{Ibanez:1991pr, Dreiner:2005rd}.
Moreover, the RPV-MSSM can explain several experimental anomalies reported in recent years including the $B$-meson anomalies~\cite{Trifinopoulos:2019lyo, Hu:2020yvs, Hu:2019ahp, Zheng:2021wnu, 
BhupalDev:2021ipu}, the ANITA anomaly~\cite{Altmannshofer:2020axr, Collins:2018jpg}, as well as the anomalous magnetic
moment of the muon~\cite{Hu:2019ahp, Zheng:2021wnu, BhupalDev:2021ipu}. If R-parity is broken, one can write down 
operators which violate baryon- or lepton-number.
Allowing all these operators to be non-vanishing would lead to a too fast proton decay rate, in conflict with the current experimental measurements~\cite{Chamoun:2020aft, ParticleDataGroup:2022pth}, unless all their couplings are extremely small.
Therefore, we assume the discrete anomaly-free baryon triality symmetry $B_3$ \cite{Dreiner:2006xw, Lee:2010vj, Dreiner:2011ft}, so that baryon-number is conserved.
In this work, we restrict ourselves to the lepton-number-violating terms only.
Further, the light bino is the lightest supersymmetric particle (LSP) in our study and decays only into SM particles via RPV couplings.

Via the lepton-number-violating RPV operators, the light bino decays lead to very similar final states as the NHL 
decays.
Moreover, the corresponding RPV couplings are all bounded to be small by various low-energy observables and
collider searches~\cite{Barbier:2004ez, Allanach:1999ic, Bolton:2021hje, Bansal:2018dge, Bansal:2019zak}. The GeV-scale
binos are hence expected to be long-lived, too, resulting in signatures such as missing energy and displaced vertices at various experiments.
These similarities raise the question: is it possible to recast the extensive exclusion bounds 
on the HNLs in the literature into corresponding bounds on the light binos in the RPV-SUSY?
In this work, we answer this question positively, by compiling a list of bounds on long-lived HNLs obtained in searches for all types of signatures 
mentioned above and recasting them into exclusion limits on the RPV-SUSY couplings as functions of the light bino mass for 
a selected list of benchmark scenarios.\footnote{Recently during the completion of this work, Ref.~\cite{Fernandez-Martinez:2023phj} appeared on arXiv; it employed similar strategies to recast the bounds on the HNLs in the minimal scenarios into those on the HNLs in effective field 
theories.} We focus on exclusion bounds acquired in past experiments, as well as predicted search sensitivities for experiments that are ongoing 
or under construction. For future (and not yet approved) experiments, we consider only 
\texttt{MoEDAL-MAPP2}~\cite{Pinfold:2019nqj, Pinfold:2019zwp} and \texttt{FASER2}~\cite{FASER:2018eoc} with 300 and 
3000\,fb$^{-1}$ integrated luminosity, respectively, as they would be the successors of some ongoing experiments at the 
LHC, while the other future concepts such as \texttt{MATHUSLA}~\cite{Curtin:2018mvb, Chou:2016lxi, MATHUSLA:2020uve}, and \texttt{ANUBIS}~\cite{Bauer:2019vqk} are independent ones and are hence not studied here. We do not consider the approved experiment \texttt{Hyper-Kamiokande}~\cite{Abe:2011ts, Hyper-KamiokandeWorkingGroup:2013hcb} for there is no 
available HNL-search sensitivity prediction that can be used with our recasting methods.

In the following section, we give the model basics of light binos in the RPV-SUSY and of the HNLs that mix with active neutrinos.
The considered experiments are introduced in Sec~\ref{sec:exp}, along with an explanation of our recasting procedure.
We then present our numerical results for some representative benchmark scenarios in Sec.~\ref{sec:results}.
Finally, in Sec.~\ref{sec:conclusions} we conclude the paper with a summary.

\section{Model basics}\label{sec:model}
\subsection{RPV-MSSM with a light bino}
In the R-parity-violating MSSM, the usual MSSM superpotential is extended by the following terms
\cite{Hall:1983id, Allanach:2003eb}:
\begin{equation}
    W_\text{RPV} = \kappa_i L_i H_u + \frac{1}{2} \lambda_{ijk} L_i L_j \bar{E}_k + \lambda'_{ijk} L_i Q_j \bar{D}_k + \frac{1}{2} \lambda''_{ijk} \bar{U}_i \bar{D}_j \bar{D}_k \, ,
    \label{eq:W_RPV}
\end{equation}
where $L_i$, $\bar E_i$, $Q_i$, $\bar U_i$, and $\bar D_i$ are chiral superfields with generation indices $i,j,k \in 
\{1,2,3\}$ and $H_u$ is one of the MSSM Higgs superfields. The $\lam_{ijk}$, $\lam'_{ijk}$ (and $\lam''_{ijk}$) are 
dimensionless Yukawa couplings, which imply lepton- (baryon-) number violating interactions and $\kappa_i$ are 
dimensionful bilinear couplings violating lepton number.
The Lagrangian in superfield-component form, as well as a complete list 
of RPV Feynman rules can be found, \textit{e.g.}, in Appendix L of Ref.~\cite{Dreiner:2008tw}. 

The RPV-MSSM allows for an unstable light long-lived neutralino, which we focus on here. A very light neutralino 
is necessarily dominantly bino-like \cite{Choudhury:1999tn} and a light bino currently avoids all experimental
and astrophysical constraints even if it is massless \cite{Gogoladze:2002xp, Dreiner:2003wh, Dreiner:2011fp}. In the 
following, we provide a concise overview of the production and decay of the lightest neutralino via $LL\bar E$ or 
$LQ\bar D$ operators. Moreover, for simplicity, we consider the lightest neutralino to be the LSP in this work. Note 
that we use the two-component fermion notation reviewed in Ref.~\cite{Dreiner:2008tw}.

\subsubsection{Neutralino production and decay via $LL\bar E$ operators}
For non-zero $LL \bar E$ couplings, neutralinos can be produced through charged lepton decays and can decay 
to lighter leptons. We provide the explicit general forms of the total decay widths for both the charged lepton and
neutralino decays, employing the matrix element and necessary phase space integration in 
Appendix ~\ref{sec:appendixA}. For all the relevant processes, we assume that all sfermions appear at the energy scale 
of the decaying particle as mass degenerate, $i.e.$, $m_{\tilde{f}} \simeq m_\text{SUSY} \gg \mchi, m_{\ell^{\pm}}$. 
Further, the R-parity conserving neutralino gauge coupling is $g'$ ($U(1)_Y$), since the neutralino is bino-like.
As a result, we can write out the coefficients appearing in the matrix elements as, (\textit{cf.}~Appendix~\ref{sec:appendixA})
\begin{equation}
    \label{eq:coeff_LLE}
  c_{ijk} \simeq -\frac{\frac{1}{\sqrt{2}} \lambda_{ijk} g' }{m_\text{SUSY}^2} \qquad \text{and} \qquad k_{ijk} \simeq \frac{\sqrt{2} \lambda_{ijk} g' }{m_\text{SUSY}^2}\, .
\end{equation}
Using Eq.~\eqref{eq:generic_LLE_width}, the relevant production widths can then be expressed as
\begin{align}
    \varGamma(\ell_k^\pm \rightarrow \tilde\chi_1^0 + \nu_i + \ell_j^\pm) = \varGamma_{LL\bar E} (\ell_k^\pm; \tilde\chi_1^0, \nu_i, \ell_j^\pm)[k_{ijk},c_{ijk},c_{ijk}] \, , \\
    \varGamma(\ell_k^\pm \rightarrow \tilde\chi_1^0 + \bar\nu_i + \ell_j^\pm) = \varGamma_{LL\bar E} (\ell_k^\pm; \tilde\chi_1^0, \bar\nu_i, \ell_j^\pm)[c_{ikj},k_{ikj},c_{ikj}] \, .
\end{align}
Similarly, the total widths of the subsequent neutralino decays $\tilde\chi_1^0 \rightarrow \overset{\scriptscriptstyle(-)}{\nu_i} + \ell_j^- + \ell_k^+$ can be written as
\begin{align}
    \varGamma(\tilde\chi_1^0 \rightarrow \nu_i + \ell_j^- + \ell_k^+) = \varGamma_{LL\bar E} (\tilde\chi_1^0; \nu_i, \ell_j^-, \ell_k^+)[c_{ijk},c_{ijk},k_{ijk}] \, , \\
    \varGamma(\tilde\chi_1^0 \rightarrow \bar\nu_i + \ell_j^- + \ell_k^+) = \varGamma_{LL\bar E} (\tilde\chi_1^0; \bar\nu_i, \ell_j^-, \ell_k^+)[c_{ikj},k_{ikj},c_{ikj}] \, .
\end{align}

\subsubsection{Neutralino production and decay via $LQ \bar D$ operators}
Via the $LQ \bar D$-operators mesons can decay into a bino accompanied by a lepton $l_i$. Subsequently, the bino 
decays via the same or another $LQ \bar D$-operator to a lepton and two quarks, where the latter again hadronize 
into a meson (for a light enough bino).
We consider charged mesons $M_{ab}^+$ with quark flavor content $(u_a \,\bar d_b)$ as well as neutral 
mesons $M_{ab}^0$ composed of $(d_a\, \bar d_b)$ and their charge conjugated equivalents. 
Neutral mesons composed of $(u_a \bar u_b)$ only contribute to higher multiplicity processes as $M \rightarrow \widetilde{\chi}^0+l_i+M^{\prime}$, where $M^{\prime}$ denotes a lighter meson and $l_i$ a charged lepton. We do \textit{not} consider them here, since these are phase space suppressed by two to three orders of magnitude~\cite{deVries:2015mfw}.
 Both the bino production and decay width are therefore given by Ref.~\cite{deVries:2015mfw}:
\begin{align}
    \varGamma(M_{ab} \rightarrow  \tilde\chi_1^0 + l_i) &= \frac{\lambda^{\frac{1}{2}}(m_{M_{ab}}^2, \mchi^2, m_{l_i}^2)}{64\pi m_{M_{ab}}^3} |G_{iab}^{S,f}|^2 (f_{M_{ab}}^S)^2\left(m_{M_{ab}}^2-\mchi^2-m_{l_i}^2\right) \, , \\
    \varGamma(\tilde\chi_1^0 \rightarrow M_{ab} + l_i) &= \frac{\lambda^{\frac{1}{2}}(\mchi^2, m_{M_{ab}}^2, m_{l_i}^2)}{128\pi \mchi^3} |G_{iab}^{S,f}|^2 (f_{M_{ab}}^S)^2\left(\mchi^2+m_{l_i}^2-m_{M_{ab}}^2\right) \, ,
\end{align}
where $\lambda^{\frac{1}{2}}(x,y,z)=\sqrt{x^2+y^2+z^2-2xy-2xz-2yz}\,$ is the square root of the K\"all\'en function and 
$l_i = \ell^{\pm}_i$ or $\nu_i$ depending on the charge of $M_{ab}$. The coefficients $|G_{iab}^{S,f}|^2$ include the 
trilinear RPV couplings and are defined in Ref.~\cite{deVries:2015mfw}, together with the meson scalar decay constants $f_{M_{ab}}^S$.

Furthermore, the $L_iL_j\bar E_j$ and $L_iQ_j\bar D_j$ operators additionally open the decay mode $\tilde
\chi_1^0 \rightarrow (\gamma + \nu_i, \gamma + \bar\nu_i)$ at the one-loop level, 
\textit{cf.}~Ref.~\cite{Dreiner:2022swd}, but this signature is not considered in the present paper since to our knowledge the corresponding HNL-decay has to-date not been searched for.

Besides the displaced-vertex signature related to the decay channels computed above, it is possible that the lightest 
neutralino is so long-lived that it does not decay inside the considered detector and appears as missing energy.

\subsection{Heavy neutral leptons}
Heavy neutral leptons are a common feature of many SM extensions attempting to give an underlying explanation
of the observed neutrino sector. The simplest HNL model one can implement is 
\begin{equation}
    \lgr \supset i \hat N^\dagger_{\alpha} \bar\sigma^\mu \partial_\mu \hat N^{\alpha} - \left[\left(Y_
    \nu\right)^i_{\alpha} \left(\varPhi^0 \hat\nu_i \hat N^\alpha - \varPhi^+\ell_i \hat N^\alpha\right) + \frac{1}{2}  
    M^\alpha_\beta \hat N_\alpha \hat N^\beta  + \text{h.c.} \right] \, ,
\end{equation}
where  $i=1,2,3$, $\varPhi^{+}$ and $\varPhi^{0}$ are the components of the SM $SU(2)_L$ Higgs doublet, and $\ell_i$ 
are the charged lepton mass eigenstates. Fields with a hat, $\hat\nu$ and $\hat N$, are the states before mass-diagonalizing the neutral lepton sector. $\left(Y_\nu\right)^i_{\alpha}$ are dimensionless Yukawa couplings and $M_\beta^\alpha = 
\text{diag}(M_{\hat N_1}, ... )$ is a diagonal mass matrix.
The index $\alpha=1,2,3,\ldots$ labels the (arbitrary many) HNLs in the theory. During electroweak symmetry breaking, 
the Higgs obtains a vacuum expectation value (vev) $v/\sqrt{2}$ with $v=246$ GeV, which gives rise to mixing of the 
HNLs $\hat N_\alpha$ with active neutrinos $\hat\nu_i$, described by the mass matrix $M_{\nu N}$:
\begin{equation}
\label{eq:hnl_massmatrix}
    M_{\nu N} = \begin{pmatrix}
    \mathbb{0}_{3\times 3} && M_D \\
    M_D^T && M
    \end{pmatrix}\, .
\end{equation}
Here, the off-diagonal entries are given by $(M_D)^i_\alpha = (Y_\nu)^i_\alpha v / \sqrt{2}$. The mass matrix can be 
perturbatively Takagi-block-diagonalized by introducing a unitary matrix $U$ \cite{Dreiner:2008tw}. For simplicity, we 
assume there is only one kinematically relevant HNL in our study. In this case, the gauge eigenstate $\hat \nu_i$ 
receives first-order contributions from the mass eigenstate $N$, proportional to the following mixing matrix entry:
\begin{equation}
    \label{eq:hnl_u}
    U_i \equiv  U^i_4 \equiv (Y_\nu^*)^{i}_1 \frac{v}{\sqrt{2} M} \, .
\end{equation}
The interaction Lagrangian with the neutrino mass eigenstates $\nu_i$ and $N$ is then given by:
\begin{equation}
    \label{eq:hnl_lgr}
    \lgr \supset -\frac{g}{\sqrt{2}} U^i_4 W^-_\mu \ell_i^\dagger \bar\sigma^\mu N - \frac{g}{2 c_W} U^i_4 Z_\mu \nu_i^\dagger \bar\sigma^\mu N + \text{h.c.} \, ,
\end{equation}
where $g$ is the $SU(2)_L$ gauge coupling and $c_W = \cos\theta_W$ is the cosine of the weak mixing angle.

\subsection{The phenomenology connecting the light bino LSP and the HNL}
The phenomenologies of the RPV-MSSM with a light bino $\tilde{\chi}_1^0$ and the SM extensions with one relevant HNL 
 turn out to be very similar. This is not surprising, as the HNL and the bino have the same gauge quantum numbers after electroweak 
symmetry breaking. 
Currently existing bounds in the HNL parameter space spanned by $(m_N, U^i_4)$ can thus be 
translated into bounds in the light-bino-RPV parameter space $(m_{\tilde\chi_1^0}, \lam/m_\text{SUSY}^2)$, where 
$\lam$ labels here any appropriate $LL \bar E$ or $LQ \bar D$ coupling.

An additional analogy between the theories can be constructed by considering the bilinear RPV couplings $\kappa_i$~\cite{Diaz:1998vf}, see Eq.~\eqref{eq:W_RPV}.
After integrating out the heavy higgsinos in the neutral fermion sector of the RPV-MSSM, one obtains a tree-level mixing of neutrinos with the bino, which is of the form
\begin{equation}
    \label{eq:nu_bino_bilinear_mixing}
    \lgr \supset \frac{g'}{2} \left( v_i -\,\frac{v_d\kappa_i}{\kappa^0} \right)\hat \nu_i \tilde\chi_1^0 + \text{h.c.}\, ,
\end{equation}
where $v_i$ and $v_d$ are the vevs of the sneutrinos and the MSSM Higgs $H_d$, respectively. $\kappa^0$ is the Higgsino mass parameter and $g'$ the $U(1)_Y$ gauge coupling.
This mixing can be interpreted as the off-diagonal entries in the neutral lepton mass matrix given in Eq.~\eqref{eq:hnl_massmatrix}, such that the elements of the matrix $U$, \textit{cf.}~Eqs.~\eqref{eq:hnl_u} and \eqref{eq:hnl_lgr}, can be mapped to the neutralino-neutrino mixing considered here:
\begin{equation}
    U^i_4 = \frac{g'}{2 \mchi} \left( v_i -\,\frac{v_d\kappa_i}{\kappa^0} \right) \, .
\label{eq:bilinearmixingmatrix}
\end{equation}
For phenomenological computations, the Lagrangian given in Eq.~\eqref{eq:hnl_lgr} can be used, which would correspond to the trivial replacement
\begin{equation}
\begin{tikzpicture}[baseline=-0.1cm]
    \setlength{\feynhandlinesize}{1.2pt}
    \begin{feynhand}
        \vertex [particle] (f1) at (-1.5, 0) {$\hat \nu$};
        \vertex [particle] (f2) at (1.5, 0) {$N$};
        \propagator [plain, insertion={[size=4pt]0.5}] (f1) to (f2);
    \end{feynhand} 
\end{tikzpicture}
\qquad\longleftrightarrow \qquad
    \begin{tikzpicture}[baseline=-0.1cm]
    \setlength{\feynhandlinesize}{1.2pt}
    \begin{feynhand}
        \vertex [particle] (f1) at (-1.5, 0) {$\hat \nu$};
        \vertex [particle] (f2) at (1.5, 0) {$\tilde\chi_1^0$};
        \propagator [plain, insertion={[size=4pt]0.5}] (f1) to (f2);
    \end{feynhand} 
\end{tikzpicture}
\end{equation}
where the inserted crosses denote the mixing. Thus, current HNL exclusion limits can be directly translated into bounds 
on the mixing strength in Eq.~\eqref{eq:nu_bino_bilinear_mixing}.
\section{Experiments and recasting}\label{sec:exp}
In this section, we present the details of existing HNL searches and classify them according to their search 
strategy. We consider experiments employing the signatures: (i) direct decays, (ii) displaced vertices, and (iii) 
missing energy. For the direct-decay searches, we further partition our analysis into: (i.a) peak searches and (i.b) 
branching ratio searches. The displaced-vertices searches are also split further into: (ii.a) beam-dump searches 
and (ii.b) collider searches.
Missing-energy searches allow us to derive new and stronger constraints on single RPV couplings. In addition, we present 
selected benchmark scenarios for which we will obtain single-coupling and coupling-product bounds within the RPV-MSSM 
framework. We provide an  overview of the experiments, discussed in this section, in Table~\ref{tab:exp_summary}.

\begin{table}[h!t]
\centering
\begin{tabular}{|c|lll|cr|}
\hline
Search Strategy & Ref. & Experiment &  Status & HNL Mixing  & HNL Mass region   \\
 \hline
\multirow{7}{*}{Peak} & \cite{PIENU:2017wbj}& \texttt{PIENU} & curr. & $|U_e|$ &  \SIrange{65}{153}{\MeV} \\ 
&\cite{PIONEER:2022alm} & \texttt{PIONEER} & proj. &  $|U_\mu|$& \SIrange{15.7}{33.8}{\MeV} \\
& \cite{PIONEER:2022alm} & \texttt{PIONEER} & proj. & $|U_e|$ & \SIrange{65}{135}{\MeV} \\
&\cite{Daum:1987bg} & \texttt{SIN} & curr. & $|U_\mu|$ & \SIrange{1}{16}{\MeV}\\
&\cite{NA62:2020mcv} & \texttt{NA62} & curr.   & $|U_\mu|$ & \SIrange{144}{462}{\MeV}\\
&\cite{NA62:2021bji} & \texttt{NA62} & curr.  & $|U_e|$ & \SIrange{200}{384}{\MeV}\\
&\cite{Asano:1981he}& \texttt{KEK} & curr.   & $|U_\mu|$ & \SIrange{160}{230}{\MeV} \\
&\cite{Hayano:1982wu} & \texttt{KEK} & curr.   & $|U_\mu|$ & \SIrange{70}{300}{\MeV}\\
\hline
\multirow{2}{*}{Branching Ratio} &\cite{PIENU:2017wbj} & \texttt{PIENU} &  curr. &$|U_e|$ &  \SIrange{0}{65}{\MeV} \\ 
& \cite{PIONEER:2022alm} & \texttt{PIONEER} & proj.  & $|U_e|$ & \SIrange{0}{65}{\MeV} \\
\hline
\multirow{9}{*}{Beam-dump} & \cite{Ballett:2019bgd} & \texttt{DUNE} & proj. &  $|U_e|$, $|U_\mu|$, $|U_\tau|$ & \SIrange{0}{1968.34}{\MeV} \\
& \cite{T2K:2019jwa} & \texttt{T2K} & curr. &   $|U_e|, |U_\mu|$ & \SIrange{10}{490}{\MeV} \\
& \cite{CHARM:1985nku,CHARMII:1994jjr} & \texttt{CHARM} & curr. &  $|U_e|$, $|U_\mu|$ & \SIrange{300}{1869.65}{\MeV} \\
& \cite{Orloff:2002de} & \texttt{CHARM} & curr. &  $|U_\tau|$ & \SIrange{290}{1600}{\MeV} \\
& \cite{NuTeV:1999kej} & \texttt{NuTeV} & curr. &  $|U_\mu|$ & \SIrange{259}{2000}{\MeV} \\
& \cite{Kelly:2021xbv} & \texttt{MicroBooNE} & curr. &  $|U_\mu|$ & \SIrange{20}{200}{\MeV}\\
& \cite{WA66:1985mfx} & \texttt{BEBC} & curr. &  $|U_e|, |U_\mu|$ & \SIrange{500}{1750}{\MeV}\\
& \cite{Barouki:2022bkt} & \texttt{BEBC} & curr. &  $ |U_\tau|$ & \SIrange{100}{1650}{\MeV}\\
& \cite{Coloma:2019htx} & \texttt{SK} & curr. &  $|U_e|$, $|U_\mu|$ & \SIrange{150}{400}{\MeV} \\
\hline
\multirow{3}{*}{Collider} & \cite{Kling:2018wct} & \texttt{FASER} & proj. & $|U_e|$, $|U_\mu|$, $|U_\tau|$ & \SIrange{0}{6274.9}{\MeV} \\
& \cite{DeVries:2020jbs} & \texttt{MoEDAL-MAPP1} & proj. &  $|U_e|$ & \SIrange{0}{6274.9}{\MeV}\\
& \cite{BaBar:2022cqj} & \texttt{BaBar} & curr. &  $|U_\tau|$ & \SIrange{100}{1360}{\MeV}\\
\hline
\multirow{2}{*}{Missing Energy} & \cite{NA62:2020pwi} & \texttt{NA62} & curr. &   $\text{BR}(\pi^0 \to \text{inv.})$ & \SIrange{0}{134.97}{MeV}\\
& \cite{BaBar:2012yut} & \texttt{BaBar} & curr. &$\text{BR}(B^0 \to \text{inv.})$ & \SIrange{0}{5279.65}{MeV}\\
\hline
\end{tabular}
\caption{Summary of experiments reviewed in Sec. \ref{sec:exp}, sorted by search strategy. We list the relevant references, the status of derived bounds (current or projected), the relevant HNL mixing, and the experimentally accessible HNL mass range.}
\label{tab:exp_summary}
\end{table}

\subsection{Direct-decay searches}

One of the main ways to produce light HNLs is via the decay of light mesons such as pions and kaons.
In direct searches, a beam of charged mesons is brought to a stop inside a scintillator where the mesons decay at rest, or 
the beam mesons are tagged and their positions, momenta, and timing information are measured by a silicon pixel spectrometer.

The energy spectrum of the visible secondary particle, \textit{i.e.}, a muon or an electron, arising from these meson 
decays is measured.  The signal shape of the energy spectrum can be compared with Monte-Carlo simulations for different 
HNL mass hypotheses. Finding no extra peaks in the secondary energy spectrum or rejecting each mass hypothesis allows us to 
exclude the relevant HNL parameters.

\subsubsection{Peak searches}
In peak searches, the energy spectrum of the secondary particle is scanned for additional peaks hinting at HNLs. 

\begin{itemize}
    \item At the 
Swiss Institute for Nuclear Research (\texttt{SIN}), a pion beam was used to put bounds on the mixing $|U_\mu|^2$ in the 
HNL mass range of \SIrange{1}{16}{\MeV}~\cite{Daum:1987bg}. 

\item A search for massive neutrinos at the \texttt{PIENU} experiment~\cite{PIENU:2017wbj} has been made in the decay of pions into positrons.
No evidence was found for additional peaks in the positron energy spectrum.
Thus, upper limits at 90\% confidence level (CL) on $|U_e|^2$ were derived in the HNL mass region \SIrange{60}{135}{\MeV}. 
In another analysis of the \texttt{PIENU} experiment~\cite{PIENU:2019usb}, heavy neutrinos were searched for in pion decays into muons.
The energy spectrum did not show any additional peaks other than the expected peak for a light neutrino.
Thus, the analysis derived a bound on $|U_\mu|^2$ for the HNL mass range of \SIrange{15.7}{33.8}{\MeV}.

\item The \texttt{PIONEER}~\cite{PIONEER:2022alm} experiment is a next-generation rare pion decay experiment. The experiment will perform the 
same search strategy as the \texttt{PIENU} experiment with higher statistics and significantly suppressed background. A 
peak search in the positron spectrum will allow probing $|U_e|^2$ in the HNL mass region \SIrange{65}{135}{\MeV}. 
Further, a search for an additional peak within the muon energy spectrum will allow us to test $|U_\mu|^2$ for $\SI{15.7}
{\MeV} < m_N < \SI{33.8}{\MeV}$.

\item \texttt{KEK}~\cite{Asano:1981he} derived an upper bound on $|U_\mu|^2$ for a massive HNL in the mass range of \SIrange{160}{230}{\MeV}.
A similar search at \texttt{KEK}~\cite{Hayano:1982wu} led to an upper bound on $|U_\mu|^2$ in the HNL mass range of \SIrange{70}{300}{\MeV}.

\item Using a kaon beam, the \texttt{NA62} collaboration placed bounds on $|U_e|^2$ for an HNL with a mass of \SIrange{144}{462}{\MeV}~\cite{NA62:2020mcv}. The analysis approach is different from \texttt{PIENU}.
In this case, a peak-search procedure measures the $K^{+} \rightarrow e^{+} N$ decay rate with respect to the $K^{+} \rightarrow e^{+} \nu$ rate for an assumed HNL mass $m_N$. The HNL mass is varied over the mentioned mass range. The benefit of this approach is the cancellations of residual detector inefficiencies, as well as trigger inefficiencies, and random veto losses.
A similar analysis~\cite{NA62:2021bji} has been performed to measure $|U_\mu|^2$ within the HNL mass range of \SIrange{200}{384}{\MeV}. 
Note that both bounds of \texttt{NA62} are derived with the assumption that the lifetime of the neutral particle exceeds \SI{50}{\ns}.

\item The \texttt{BaBar} experiment~\cite{BaBar:2022cqj}  at SLAC has performed a search for the rare decay  
$\tau^- \to \pi^- \pi^- \pi^+ + N$ in the mass region of $100 < m_N < \SI{1360}{\MeV}$. The observed kinematic phase 
space distribution of the hadronic system allows \texttt{BaBar} to place a stringent bound on $|U_\tau|^2$. However, the 
search is based on three-prong tau events. Technically, this allows us to derive a single coupling bound on $\lambda^
\prime_{311}$. However, the four-body production mode of the light neutralino would need proper phase-space 
consideration. Therefore, we do not include a reinterpretation of this search in our work and shall discuss it
elsewhere.

\end{itemize}

\subsubsection{Branching-ratio searches}
It is possible to measure branching ratios of different pion decay modes. The ratio 
\begin{align}
R_{e/\mu} = \frac{\Gamma\left(\pi^+ \to e^+ + \nu(\gamma)\right)}{\Gamma\left(\pi^+ \to \mu^+ + \nu(\gamma)\right)},
\end{align}
can be used to derive limits on the mixing $|U_e|^2$ in the region $m_N<\SI{65}{\MeV}$.
This has been performed by \texttt{PIENU}~\cite{PiENu:2015seu} and is planned for \texttt{PIONEER}~\cite{PIONEER:2022alm}.

\subsection{Displaced-vertex searches}
Beam-dump and collider experiments can produce HNLs via the same processes that produce light neutrinos. A proton beam 
hitting a fixed target typically produces a large number of pions and kaons, and also heavier mesons. If kinematically 
allowed, the decay of the primary mesons can produce HNLs, which will propagate freely since they are long-lived and 
interact only feebly. The HNLs produced at beam-dump experiments are typically boosted in the forward direction, which 
further increases their decay length in the lab frame. Hence, only a fraction of the produced HNLs decay at the location 
of the detector. To reduce possible background events, the experiments usually have a system of veto detectors equipped 
for both charged and neutral particles. The search strategy relies on the visibility of the HNL decay products inside 
the detector; we will discuss these later.

In the following, we present displaced-vertex searches at beam-dump and collider 
experiments separately.

\subsubsection{Beam-dump search}
\begin{itemize} 
\item At \texttt{DUNE}~\cite{DUNE:2020lwj}, HNLs can be produced via pion-, kaon-, and  
$D$-meson-decays.
Ref.~\cite{Ballett:2019bgd} predicts constraints on HNLs by searching for their decay products inside the \texttt{DUNE} Near Detector.
It is assumed that only the three-neutrino final state is not detectable.
The search strategy allows to measure all three mixings $|U_e|^2, |U_\mu|^2$ and $|U_\tau|^2$ for an HNL mass range up to the mass of the $D_s$-meson.
Note that in the analysis, a single mixing element is assumed to dominate over the other two at a time.

\item The \texttt{T2K} experiment~\cite{T2K:2019jwa} follows the same approach but uses a kaon beam.
Thus, they derive bounds on both $|U_e|^2$ and $|U_\mu|^2$ for $\SI{10}{\MeV} < m_N < \SI{490}{\MeV}$.

\item Heavy neutral leptons in the \texttt{CHARM} beam-dump experiment are produced from $D$ and 
$D_s$-meson decays.\footnote{In this work, we use ``$D$-mesons'' (``$B$-mesons'') to label the $D^\pm$ 
mesons ($\overset{\scriptscriptstyle(-)}{B^0}$ and $B^\pm$ mesons), while $D_s$ and $B_c$ mesons are
separately discussed. We do not take into account $\overset{\scriptscriptstyle(-)}{D^0}$ or $\overset 
{\scriptscriptstyle(-)}{B_s}$ mesons.} The former allows to set bounds on $|U_e|^2, |U_\mu|^2$ 
for an HNL mass of \SIrange{300}{1869.65}{\MeV}~\cite{CHARM:1985nku,CHARMII:1994jjr}, and the latter 
allows to probe $|U_\tau|^2$ for $\SI{290}{\MeV} < m_N < \SI{1600}{\MeV}$~\cite{Orloff:2002de}.

\item A search for HNLs has been performed at the \texttt{NuTeV} experiment~\cite{NuTeV:1999kej} at 
Fermilab. The data were examined for HNLs decaying into muonic final states to derive bounds on the 
mixing $|U_\mu|^2$ of HNLs in the \SIrange{0.25}{2.0}{\GeV} mass range.
See also Ref.~\cite{Dedes:2001zia} which directly considers the \texttt{NuTeV} data in terms of a light neutralino.

\item An analysis of current data from the \texttt{MicroBooNE} experiment~\cite{Kelly:2021xbv} can 
constrain the parameters of HNLs, that mix predominantly with muon-flavored 
neutrinos, for HNL masses between \SIrange{20}
{200}{\MeV}.

\item The \texttt{BEBC} experiment derived limits on the HNL-light neutrino mixing parameters from a search for decays of heavy neutrinos in a proton beam-dump 
experiment~\cite{WA66:1985mfx}. It derived bounds on $|U_e|^2, |U_\mu|^2$ for an HNL mass between 
\SI{0.5}{\GeV} and \SI{1.75}{\GeV}. A re-analysis~\cite{Marocco:2020dqu, Barouki:2022bkt} has 
demonstrated that the \texttt{BEBC} detector was also able to place bounds on the $|U_\tau|^2$ 
mixing for HNL masses higher than the kaon mass. This re-analysis has taken into account several 
production and decay channels of HNLs.

\item For \texttt{Super-Kamiokande (SK)}, the largest contribution to HNL production is 
through the decay of mesons produced in the atmosphere via cosmic rays.
A secondary contribution to the flux comes from the HNL production in neutral-current scattering of atmospheric neutrinos passing through the Earth.
The total number of HNL decays inside the detector within a given time window results in an upper bound on the HNL mixing. This approach is different from the displaced-decay search limits from beam dumps and allows one to derive bounds on $|U_e|^2$ and $|U_\mu|^2$ in the minimal HNL scenarios for masses between \SIrange{150}{400}{\MeV}~\cite{Coloma:2019htx}.

\end{itemize}

\subsubsection{Collider searches}
HNLs can be also searched for in (semi-)leptonic decays of mesons produced at the LHC via a similar mechanism 
as in beam-dump experiments.
At colliders, particle collisions produce an abundance of mesons, such as pions, kaons, $D$-mesons, and $B$-mesons, which can further decay into HNLs.
Far detectors at the LHC are sensitive to decaying, light LLPs with a decay length  comparable
with their distance to the interaction point, \textit{e.g.},  of $\mathcal{O}(1)-\mathcal{O}(100)$ \SI{}{\m}. 
HNLs decaying inside the detector can be identified by their decay products, except for the 
invisible three-neutrino final state.
To ensure a low background environment, the experiments are proposed to be set far away from the primary proton-proton collision points and require shielding between the interaction points and the detectors.

Some of the new detectors at the LHC are already approved and currently running: \texttt{FASER}~\cite{Feng:2017uoz} 
and \texttt{MoEDAL-MAPP1}~\cite{Pinfold:2019nqj}. We include these running HNL searches in order to 
reinterpret their projected sensitivity.
Their follow-up programs, \texttt{FASER2}~\cite{FASER:2018eoc} and \linebreak {\texttt{MoEDAL-MAPP2}~\cite{Pinfold:2019zwp}}, have been proposed for operation during the high-luminosity LHC phase, with an expected final integrated luminosity \SI{3}{\per\atto\barn} and \SI{300}{\per\femto\barn} respectively, and are taken into account in our numerical analysis, as well.

\texttt{FASER(2)}~\cite{Kling:2018wct}\footnote{In principle, Ref.~\cite{FASER:2018eoc} employs the most 
updated geometrical setup of \texttt{FASER(2)}, but the results shown therein for the HNLs do not separate 
the contributions from $D$- and $B$-mesons. Therefore, we have chosen to reinterpret the results given in 
Ref.~\cite{Kling:2018wct} for the HNLs from the heavy mesons' decays, which are only slightly different 
from those given in Ref.~\cite{FASER:2018eoc}.} is intended to detect long-lived 
particles decaying inside the detector volume.
A sensitivity estimate, taking the detector geometry into account, leads to a specific reach in $|U_e|^2, |U_\mu|^2$, and $|U_\tau|^2$ for an HNL with a mass up to the mass of the $B_c$-meson.
Similarly, \texttt{MoEDAL-MAPP1(2)}~\cite{DeVries:2020jbs,Beltran:2023nli} can probe the mixing $|U_e|^2$ for HNL being produced in decays of $D$- and $B$-mesons for the same mass range as \texttt{FASER(2)}.

In addition, other experimental proposals for LLP far detectors include \texttt{MATHUSLA}~\cite{Chou:2016lxi, 
Curtin:2018mvb, MATHUSLA:2020uve}, \texttt{ANUBIS}~\cite{Bauer:2019vqk}, \texttt{CODEX-b}~\cite{Gligorov:2017nwh}, 
and \texttt{FACET}~\cite{Cerci:2021nlb}. None of these proposed detectors has been officially approved. We 
therefore do \textit{not} include them in our reinterpretation strategy. We still want to 
emphasize the prospect of these experiments in the search for light long-lived particles. Sensitivity estimates, 
worked out in detail on the minimal HNL scenarios in Ref.~\cite{DeVries:2020jbs}, would also provide possible 
discovery potential for the discussed light neutralino scenarios, as worked out for example in 
Refs.~\cite{Dercks:2018eua,Dercks:2018wum, Dreiner:2020qbi,Dreiner:2022swd}. These potential future experiments all intend to 
look for various BSM signatures which include neutralino decays induced by all of the $LH_u$, $LL \bar E$, 
and $LQ\Bar{D}$ operators.

\medskip

\medskip

\subsection{Missing-energy searches}

The \texttt{NA62} search~\cite{NA62:2020pwi} allows us to derive bounds on the branching ratio of pions 
decaying into an invisible final state. This is achieved by the reconstruction of the charged particles 
in the process $K^+ \to \pi^+ \pi^0$. The analysis relies on the tracking of the charged $K^+$ and $\pi
^+$ and can probe $\pi^0$ decay to any invisible final state. \texttt{NA62} 
reported a 90\% CL upper limit on $\text{BR}(\pi^0 \to \text{inv.}) < \SI{4.4e-9}
{}$. The search can be recast to derive bounds on RPV couplings which also contribute to the invisible 
decays of the pion. It turns out that the obtained limits are weaker than the currently existing bounds~\cite{Allanach:1999ic, Dercks:2017lfq} and are, therefore, omitted.

The \texttt{BaBar} experiment has also searched for rare decays $B_0 \to \text{inv.}$~\cite{BaBar:2012yut}.
The search relies on the identification of the other neutral $B$-meson, as the ‘tag side’, and thus, can 
measure a purely invisible decay width of the $B$-meson.
The upper limit at the 90\% CL yields $\text{BR}(B_0 \to \text{inv.}) < \SI{2.4e-5}{}$.
We can recast this search into bounds on the production coupling of a light long-lived neutralino.

We could not find existing searches for invisible decays of the other uncharged 
mesons, \textit{i.e.}, $K_S$, $K_L$, and $B_s$. To derive limits on the decay of these mesons into a 
neutralino we consider the uncertainty of the total decay width of the mesons. We assume all visible 
decay modes to be measured within the decay width. The resulting 
uncertainty can be extended to account for additional invisible decays and potential errors in 
measurements. Thus, we use the uncertainty to establish an upper limit on the branching ratio into 
invisible final states, if kinematically allowed. These could contain 
neutrinos, HNLs, or neutralinos. Assuming the latter saturates the width uncertainty, we derive bounds on 
the $LQ\Bar{D}$ couplings. The measured uncertainties can be found in Ref.~\cite{ParticleDataGroup:2022pth}.

We could not find bounds from direct searches for massive HNLs in the leptonic 
decays $\mu^-\to e^-+\nu  + N$ and $\tau^- \to e^-/\mu^-+ \nu + N$. There are only searches for the muons 
and the $\tau$ leptons to decay to a lighter charged lepton plus active neutrinos or a 
photon~\cite{MEG:2016leq}. Thus, the existing limits on the branching ratio cannot be generalized for a 
massive HNL owing to kinematic assumptions. In this case, we again rely on the uncertainty of the decay widths. This allows us to derive bounds on the $LL
\bar{E}$ couplings. Again, the uncertainties of the decay width are taken from 
Ref.~\cite{ParticleDataGroup:2022pth}.

{\color{white}.

.}

\subsection{Other searches}
For the bilinear coupling scenarios, we use the most relevant and up-to-date constraints on the HNLs existing in the literature, which are summarized in Ref.~\cite{Fernandez-Martinez:2023phj}.
Therefore, we supplement the results from the previously discussed experiments with data from \texttt{TRIUMF}~\cite{Britton:1992xv}, 
\texttt{PSI}~\cite{Daum:1987bg}, \texttt{Borexino}~\cite{Borexino:2013bot,Plestid:2020ssy}, and atmospheric 
neutrinos scattering in the Earth~\cite{Dentler:2018sju}.

\subsection{The recasting procedure}
The procedures for recasting the HNL bounds into limits on RPV scenarios depend on the RPV couplings that 
are switched on, and the search strategy of the experiments; they fall into one of the following three 
categories: 
\begin{itemize}
    \item The most straightforward case is for scenarios involving bilinear RPV couplings. As discussed in 
    Sec.~\ref{sec:model}, these couplings lead to mixing between the neutralino and the neutrinos, 
    \textit{cf}.~Eq.~\eqref{eq:nu_bino_bilinear_mixing}. Thus, we directly translate HNL exclusion limits 
    in the mixing vs.~mass plane into bounds in the RPV coupling vs.~neutralino mass plane, using 
    Eq.~\eqref{eq:bilinearmixingmatrix}.
    \item In RPV scenarios involving $LQ\Bar{D}$ or $LL\Bar{E}$ operators, and a (detector-level) stable 
    neutralino produced in the decay of a meson or lepton, missing-energy searches and peak searches can 
    provide sensitivity. We use the HNL exclusion limits (typically in the mixing vs.~mass plane) in order 
    to determine the bounds on the decay width of the relevant meson/lepton into an HNL (see, for instance, 
    Refs.~\cite{Gorbunov:2007ak,Gorbunov:2020rjx}). Since replacing the HNL with a light bino of the same 
    mass does not change the experimental signature and the kinematics, we can simply equate the above with 
    the corresponding decay width of the same meson/lepton into a bino in the RPV model, using the 
    expressions given in Sec.~\ref{sec:model}; this gives us the bound on the relevant RPV coupling in 
    terms of the bino mass.
    \item For RPV scenarios involving $LQ\Bar{D}$ or $LL\Bar{E}$ operators where the neutralino is produced
    in meson/lepton decays, and is \textit{unstable} at the detector scales, displaced-vertex searches (beam dump or collider) can apply. In this case, we use the long-lifetime approximation to calculate the number 
    of decay events reconstructed in the detector, following the arguments outlined in 
    Ref.~\cite{Beltran:2023nli} (see also Ref.~\cite{Fernandez-Martinez:2023phj}); we review the procedure briefly now. Consider a beam-dump or collider 
    experiment that searches for HNLs via displaced vertices. Let the HNL, $N$, be produced in the decay of 
    a parent particle $P$ (for instance, a pion).\footnote{We are assuming that the direct production of 
    the HNL (or bino) is suppressed compared to indirect production via decays of mesons and leptons, which is 
    typically the case.} Then, the number of detected events for a final state, $Y$, produced in the decay of $N$ can be estimated as,
    \begin{equation}
        N^{\text{HNL}}_{\text{events}} = N_P\times \text{BR}(P\to N+X) \times \text{BR}(N\to Y) \times \epsilon\,,
        \label{eq:HNL_events}
    \end{equation}
    where $N_P$ is the number of $P$ produced at the experiment, $X$ are any additional objects produced in 
    $P$'s decay that are not of interest to us, and the BRs are the corresponding branching ratios. $\epsilon$ is a factor accounting for the detector acceptance and efficiency. This 
    factor is linearly proportional to the probability for the HNL to decay inside the detector ($P[
    \text{decay}]$). It also depends on experiment-specific information such as the detector type and its geometry.
    In the limit of a long-lived HNL, such that the boosted decay length is much 
    larger than the distance $\Delta L$ from the interaction point of the experiment to the first edge of the 
    detector along the direction of travel of the HNL, to a good approximation we have~\cite{Beltran:2023nli},
    \begin{equation}
        P[\text{decay}] \approx \frac{\Delta L \times \Gamma_N}{\beta_N\gamma_N}\,,
    \end{equation}
    where $\Gamma_N, \beta_N$, and $\gamma_N$ represent the total decay width, the relativistic velocity, 
    and the Lorentz boost factor of the HNL, respectively. Inserting this in Eq.~\eqref{eq:HNL_events}, 
    we obtain,
    \begin{equation}
        N^{\text{HNL}}_{\text{events}} = A\times\frac{\text{BR}(P\to N+X) \times \Gamma(N\to Y)}{\beta_N\gamma_N}\,,
        \label{eq:HNL_events_simpler}
    \end{equation}
    where all the quantities depending on the HNL model details are written explicitly, and the remaining 
    experiment-specific factors of Eq.~\eqref{eq:HNL_events_simpler} have been absorbed into the 
    proportionality factor $A$. Now, assume that there is also an RPV scenario in which the light bino, $\tilde
    {\chi}^0_1$, is produced in $P$ decays, and has a decay mode into $Y$. Analogously, we can 
    write,
    \begin{equation}
        N^{\text{RPV}}_{\text{events}} = A\times\frac{\text{BR}(P\to \tilde{\chi}^0_1+X') \times \Gamma(\tilde{\chi}_1^0\to Y)}{\beta_{\tilde{\chi}_1^0}\gamma_{\tilde{\chi}_1^0}}\,,
        \label{eq:RPV_events}
    \end{equation}
    where the HNL is now replaced by the bino. Here, since both the light bino and the HNL are produced in the same meson's decay at the identical experiment, the proportionality factor $A$ can 
    be legitimately assumed to be essentially the same. Thus, combining the two equations, we obtain a simple 
    scaling relation between the two models:
    \begin{equation}
        \frac{N^{\text{RPV}}_{\text{events}}}{N^{\text{HNL}}_{\text{events}}} = \frac{\text{BR}(P\to \tilde{\chi}^0_1+X') \times \Gamma(\tilde{\chi}^0_1\to Y) \times \beta_N\gamma_N}{\text{BR}(P\to N+X) \times \Gamma(N\to Y) \times \beta_{\tilde{\chi}^0_1}\gamma_{\tilde{\chi}^0_1}}\,.
        \label{eq:HNL_RPV_events}
    \end{equation}
    For an HNL and a light bino with the same mass, we can further simplify the above expression in the limit where the HNL and the bino carry the same energy in the lab frame.
    In this case, the $\beta \gamma$ factors cancel out, and we have,
    \begin{equation}
        \frac{N^{\text{RPV}}_{\text{events}}}{N^{\text{HNL}}_{\text{events}}} = \frac{\text{BR}(P\to \tilde{\chi}^0_1+X') \times \Gamma(\tilde{\chi}^0_1\to Y)}{\text{BR}(P\to N+X) \times \Gamma(N\to Y)}\,,
        \label{eq:HNL_RPV_events_simpler}
    \end{equation}
    which is completely free from any experiment-specific factors. This will be the master expression we use for recasting DV searches.
    We note that, technically, having different $X$ and $X'$ induces different kinematics for the HNL and the bino.
    For instance, the two final states could have different masses, or even contain differing number of objects, thus affecting the energy carried by the bino relative to the HNL.
    However, typically, the experiments we consider produce the parent particles (or their decay products) with significant boosts, in which case the above formula is only modified mildly.
    On the other hand, we stress that the HNL and the bino must be produced in the decay of the same (or similar) parents, and must decay into the same final state since these can significantly alter the detector acceptances and efficiencies.\footnote{Actually, the final states into which the HNL and bino decay need not be identical, \textit{e.g.}, the invisible objects contained in both do not need to match; the crucial part is that the detection efficiencies at the considered experiment must be the same. } 
    
    To use Eq.~\eqref{eq:HNL_RPV_events_simpler}, assume that a given DV search for HNLs concludes without discovery, and obtains bounds on the HNL model parameters corresponding to a certain signal-event number, $[N_{\text{events}}^{\text{HNL}}]_{\text{bound}}$.
    Since the signal (and kinematics) in the RPV model is the same, this bound also applies to the light binos: $[N_{\text{events}}^{\text{RPV}}]_{\text{bound}} = [N_{\text{events}}^{\text{HNL}}]_{\text{bound}}$. Plugging this into Eq.~\eqref{eq:HNL_RPV_events_simpler}, we arrive at,
    \begin{align}
        [\text{BR}(P\to \tilde{\chi}^0_1+X') \times \Gamma(\tilde{\chi}^0_1\to Y)]_{\text{bound}}=[\text{BR}(P\to N+X) \times \Gamma(N\to Y)]_{\text{bound}}\,.
    \end{align}
    The right-hand side of the above equation can be evaluated by using the bounds on the HNL mass and mixings as input (see, for instance, Refs.~\cite{Shrock:1980ct, Gorbunov:2007ak, Bondarenko:2018ptm} for explicit expressions for all relevant decay widths in terms of these parameters), while the expressions for the RPV counterparts appear in Sec.~\ref{sec:model} and depend on the RPV couplings, sfermion masses, as well as the neutralino mass.
\end{itemize}

\section{Numerical results}\label{sec:results}
In order to present our results, we consider benchmark scenarios with one or two non-vanishing RPV 
couplings. Additional non-zero RPV couplings could allow for further neutralino decay channels to 
open up; these would modify the relevant branching ratios and neutralino decay length, and hence, 
the presented sensitivity limits.

\subsection{One-coupling scenarios}
We first consider scenarios where only one non-zero RPV operator contributes to the relevant 
physical process at a time. For each of the RPV couplings in Eq.~\eqref{eq:W_RPV} (except the $\lam''
_{ijk}$'s), we list the relevant HNL searches providing constraints in~\cref{tab:single_couplings_BMs}.
We now discuss the sensitivity limits for each category of RPV coupling in detail.

\begin{table}[!ht]
\centering
\begin{adjustbox}{width=\columnwidth}
\begin{tabular}{|c|c|c|c|c|}
\hline
\bf{Coupling} & \bf{Direct Decays} & $\bm{E_T^\text{miss}}$ & \bf{DV} & \bf{Label}\\
\hline
$\kappa^i$ & $\nu$-mixing~\cite{Fernandez-Martinez:2023phj} & $\nu$-mixing~\cite{Fernandez-Martinez:2023phj} & $\nu$-mixing~\cite{Fernandez-Martinez:2023phj} & $\bm{\kappa}$\\
\hline
$\lambda'_{a11}$ & $\pi^\pm\to e^\pm_a + \tilde{\chi}^0_1$
& $\pi^0\to \text{invis.}$~\cite{NA62:2020pwi} & $\bm{\times}$ & \multirow{4}{*}{$\bm{\lambda'_{\pi}}$}\\[0.13 cm]
&~\cite{PiENu:2015seu, PIENU:2017wbj, PIENU:2019usb, PIONEER:2022alm} &&& \\[0.15 cm]
$\lambda'_{311}$ & $\left(\bm{\times}\right) \tau^\pm\to \{\pi^\pm/\rho^\pm\} + \tilde{\chi}^0_1$ & $\pi^0\to \text{invis.}$~\cite{NA62:2020pwi} & $\tau^\pm\to \pi^\pm + \tilde{\chi}^0_1; \tilde{\chi}^0_1 \to \nu_{\tau} + \{\pi^0/\rho^0/\eta/\eta'/\omega\}$
& \\[0.15 cm]
&&&\cite{Ballett:2019bgd, Kling:2018wct, DeVries:2020jbs, Beltran:2023nli}&\\
\hline
$\lambda'_{a12}$ & $K^\pm\to e^\pm_a + \tilde{\chi}^0_1$ & $K^0_L\to \text{invis.}$~\cite{ParticleDataGroup:2022pth} & $\bm{\times}$ & \multirow{4}{*}{$\bm{\lambda'_{K}}$}\\[0.15 cm]
&~\cite{Asano:1981he, Hayano:1982wu, NA62:2020mcv, NA62:2021bji} &&&\\[0.15 cm]
$\lambda'_{312}$ & $\left(\bm{\times}\right) \tau^\pm\to K^\pm + \tilde{\chi}^0_1$ & $K^0_L\to \text{invis.}$~\cite{ParticleDataGroup:2022pth} & $\bm{\times}$ & \\[0.15 cm]
\hline
$\lambda'_{i13}$ & $\left(\bm{\times}\right) B^\pm\to e^\pm_i + \tilde{\chi}^0_1$ & $B^0\to \text{invis.}$~\cite{BaBar:2012yut} & $\bm{\times}$ & $\bm{\lambda'_{B1}}$\\[0.15 cm]
\hline
$\lambda'_{i21}$ & $\left(\bm{\times}\right) D^\pm\to e^\pm_i + \tilde{\chi}^0_1$ & $K^0_L\to \text{invis.}$~\cite{ParticleDataGroup:2022pth} & $\bm{\times}$ & $\bm{\lambda'_{D/K}}$\\[0.15 cm]
\hline
$\lambda'_{i22}$ & $\left(\bm{\times}\right) D_s^\pm\to e^\pm_i + \tilde{\chi}^0_1$ & $\bm{\times}$ & $D_s^\pm\to e^\pm_i + \tilde{\chi}^0_1; \tilde{\chi}^0_1\to \nu_e + \{\phi/\eta/\eta'\}$ 
& $\bm{\lambda'_{D_s}}$\\[0.15 cm]
&&& \cite{Ballett:2019bgd, Kling:2018wct, DeVries:2020jbs, Beltran:2023nli} &\\[0.15cm]
\hline
$\lambda'_{i23}$ & $\left(\bm{\times}\right) B_c^\pm\to e^\pm_i + \tilde{\chi}^0_1$ & $B_s^0\to \text{invis.}$~\cite{ParticleDataGroup:2022pth} & $\bm{\times}$ & $\bm{\lambda'_{B_c/B_s}}$\\[0.15 cm]
\hline
$\lambda'_{i31}$ & $\bm{\times}$ & $B^0\to \text{invis.}$~\cite{BaBar:2012yut} & $\bm{\times}$ & $\bm{\lambda'_{B2}}$\\[0.15 cm]
\hline
$\lambda'_{i32}$ & $\bm{\times}$ & $B^0_s\to \text{invis.}$~\cite{ParticleDataGroup:2022pth} & $\bm{\times}$ & $\bm{\lambda'_{B_s}}$\\[0.15 cm]
\hline
$\lambda'_{i33}$ & $\bm{\times}$ & $\bm{\times}$ & $\bm{\times}$ & $\bm{-}$\\[0.15 cm]
\hline
$\lambda_{12a}$ & $\bm{\times}$ & $\mu^\pm \to e^\pm + \text{invis.}$
& $\bm{\times}$ & $\bm{\lambda_{\mu}}$\\[0.15 cm]
&&\cite{ParticleDataGroup:2022pth}&& \\
\hline
$\lambda_{123}; \lambda_{13i};$ & \multirow{2}{*}{$\bm{\times}$} & \multirow{2}{*}{$\tau^\pm \to \{e^\pm/\mu^\pm\} + \text{invis.}$}
& \multirow{2}{*}{$\bm{\times}$} & \multirow{2}{*}{$\bm{\lambda_{\tau}}$}\\[0.21 cm]
$\lambda_{232}; \lambda_{233}$ & & \cite{ParticleDataGroup:2022pth} & &\\[0.15 cm]
\hline
$\lambda_{231}$ & $\bm{\times}$ & $\{\tau^\pm/\mu^\pm\} \to e^\pm + \text{invis.}$
& $\bm{\times}$ & $\bm{\lambda_{\tau/\mu}}$\\[0.15 cm]
&&\cite{ParticleDataGroup:2022pth}&& \\
\hline
\end{tabular}
\end{adjustbox}
\caption{Details of the searches providing constraints when only one non-zero RPV operator contributes
at a time. We list all the bilinear, $LQ\Bar{D}$, and $LL\Bar{E}$ operators in the first column (by coupling).
The second to fourth columns contain the physical processes that provide constraints and the references to the relevant 
existing HNL searches targeting them. The fifth column indicates our labeling scheme for the scenarios.
$\bm{\times}$ denotes the absence of a constraining process, while $\left(\bm{\times}\right)$ labels that, 
in principle, the listed process may provide constraints but we could not find a relevant existing HNL search.
In the table, $a\in\{1,2\}$ and $i\in\{1,2,3\}$.}
\label{tab:single_couplings_BMs}
\end{table}

\subsubsection{Bilinear scenarios}
\begin{figure}[!ht]
	\centering
	\includegraphics[width=0.8\textwidth]{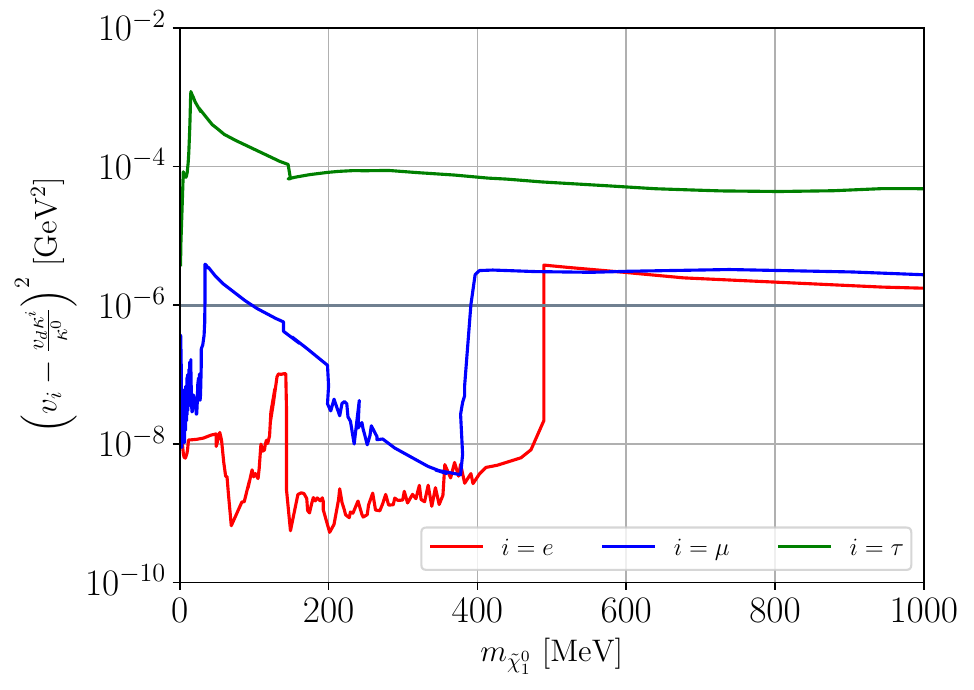}
	\caption{Exclusion limits on bilinear RPV couplings as a function of the light bino mass, reinterpreted from existing HNL searches. The current limit on the bilinear couplings is shown as a horizontal gray line.}
	\label{fig:bilinear_bounds}
\end{figure}

The bilinear couplings ($\kappa^i$ and the sneutrino vacuum expectation values) induce a mixing between 
the light bino and the three light neutrinos, \textit{cf}.~Eq.~\eqref{eq:nu_bino_bilinear_mixing}. Thus, 
all HNL searches constraining the mixing between the HNL and the neutrinos directly imply constraints on 
these couplings. The exclusion limits are shown in Fig.~\ref{fig:bilinear_bounds}. These have been read 
off from Ref.~\cite{Fernandez-Martinez:2023phj}. The bilinear couplings also generate mass terms for the 
neutrinos -- see, for instance, Refs.~\cite{Hall:1983id, Joshipura:1994ib, Nowakowski:1995dx, Banks:1995by, 
Allanach:2003eb} -- leading to the constraint $v_i,\,\kappa^i \lesssim \mathcal{O}\left(\SI{1}{\mega
\electronvolt}\right)$~\cite{Allanach:2003eb}; this is depicted as a gray horizontal line in the 
plot.\footnote{We note that the existing limit comes from the cosmological bound on the 
neutrino masses, and is thus scenario-dependent~\cite{ParticleDataGroup:2022pth}. Further, there is also dependence 
on undetermined supersymmetric parameters.} We find that the reinterpreted bounds from the HNL searches 
corresponding to the light-flavor neutrinos ($e$ and $\mu$) have sensitivity to regions beyond the current limits at low neutralino masses below 500 MeV.
In this region, the most constraining limits come from \texttt{PIENU}~\cite{PiENu:2015seu, PIENU:2017wbj,PIENU:2019usb}, \texttt{NA62}~\cite{NA62:2020mcv, NA62:2021bji}, and \texttt{T2K}~\cite{T2K:2019jwa, Arguelles:2021dqn}.
Beyond this mass, and for the tau case -- where charged kaon and pion decay into HNLs are kinematically forbidden -- existing limits are weaker.

\subsubsection{$LQ\bar{D}$ scenarios} 
\begin{figure}[!ht]
\begin{subfigure}{0.49\textwidth}
  \includegraphics[width=\linewidth]{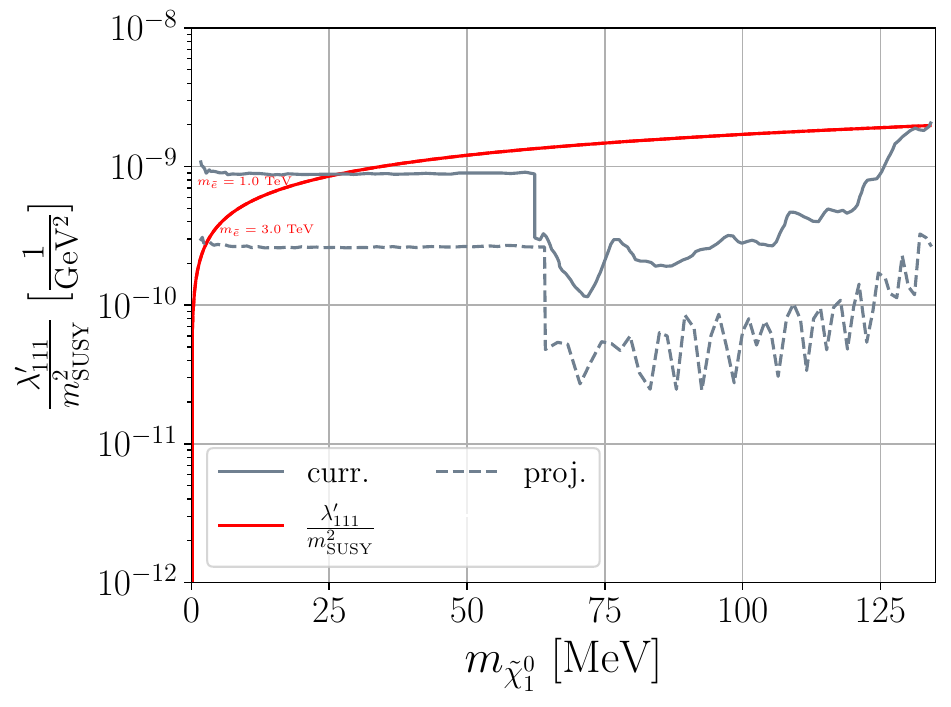}
  \caption{$\lambda'_{111}$.}
  \label{fig:pion1}
  \end{subfigure}\hfill
\begin{subfigure}{0.49\textwidth}%
    \includegraphics[width=\linewidth]{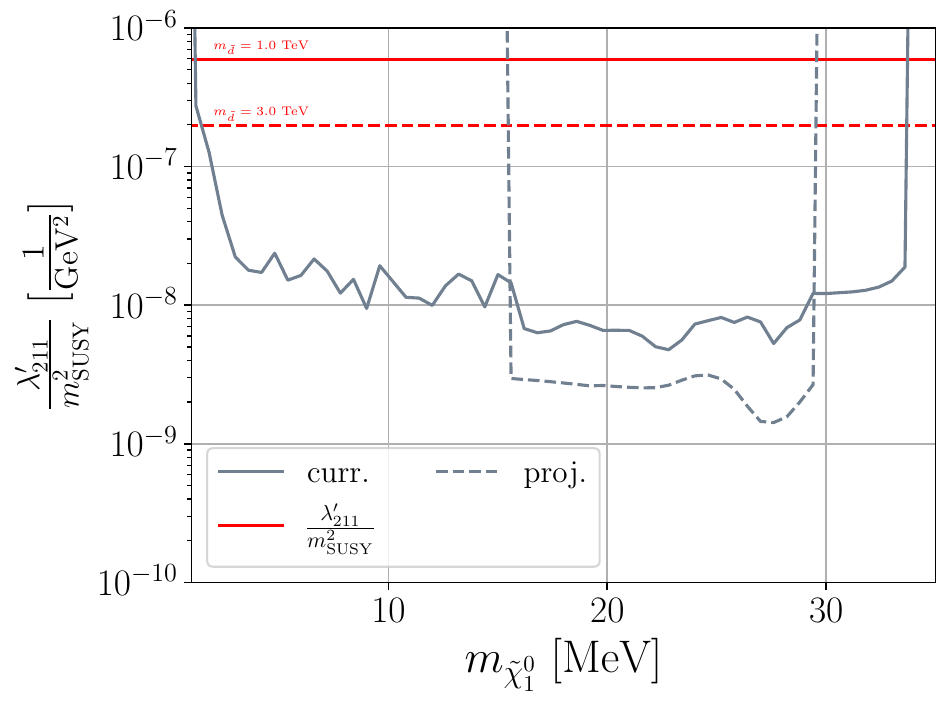}
  \caption{$\lambda'_{211}$.}
  \label{fig:pion2}
\end{subfigure}
\\
\centering
\begin{subfigure}{0.49\textwidth}
  \includegraphics[width=\linewidth]{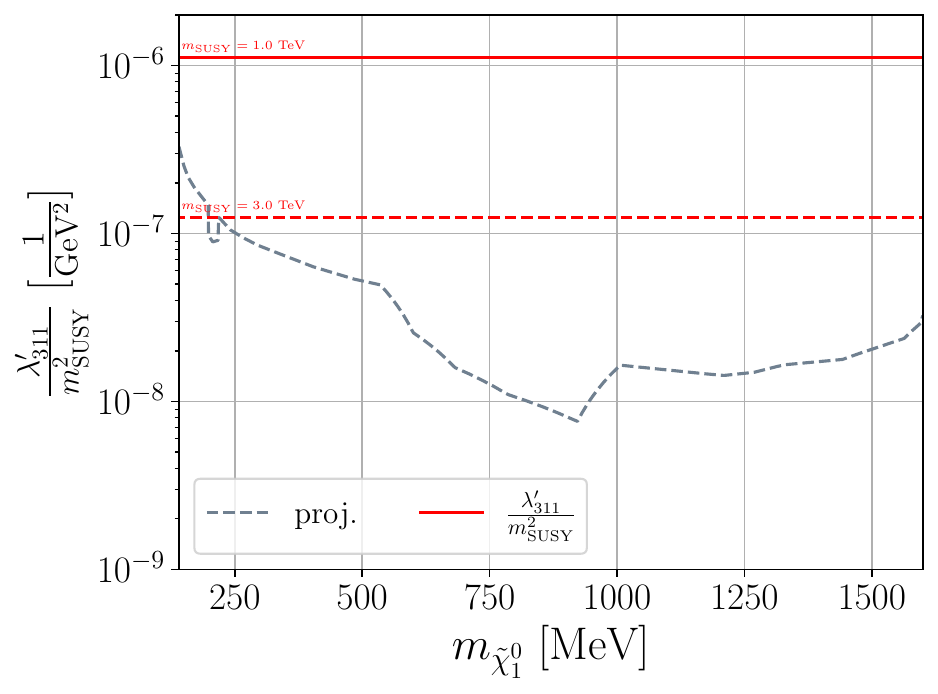}
  \caption{$\lambda'_{311}$.
  }
  \label{fig:pion3}
  \end{subfigure}
\caption{Sensitivity limits on the $\bm{\lam'_{\pi}}$ one-coupling scenarios 
of~\cref{tab:single_couplings_BMs} as a function of the light bino mass, reinterpreted from 
existing HNL searches. Current (projected) limits obtained from the reinterpretation are 
shown as solid (dashed) gray lines. The existing limits on the RPV couplings are also shown 
in red, with the solid and dashed lines corresponding to varying assumptions of unknown SUSY mass scales.}
\label{fig:pion_single}
\end{figure}

For the remaining one-coupling scenarios, all existing limits on the RPV couplings are taken from 
Ref.~\cite{Dercks:2017lfq}. The $L_iQ_1\bar D_1$-operators couple to pions. With $i=1$, the decay $\pi^\pm 
\to e^\pm +\tilde{\chi}^0_1$ is allowed, for masses $m_{\tilde{\chi}^0_1} \le m_{\pi^{\pm}}-m_{e}$.\footnote{In
the small region of phase space, $m_{\pi^0}+m_{\nu_e} < m_{\tilde{\chi}^0_1} < m_\pi^\pm-m_e$, $\lam
'_{111}\not=0$ allows for the neutralino to decay into a neutrino and a pion; displaced-vertex 
searches can constrain this process. However, we have ignored this in~\cref{tab:single_couplings_BMs}
(also analogously for the kaon).} This process has been searched for in the context of HNLs, and the
most stringent existing limits are provided by \texttt{PIENU}~\cite{PIENU:2017wbj}. Additionally, 
the approved \texttt{PIONEER}~\cite{PIONEER:2022alm} experiment is projected to have sensitivity 
beyond these limits. We show the resulting contours in~\cref{fig:pion1}. The sharp drop in sensitivity 
at $m_{\tilde{\chi}^0_1} \approx \SI{65}{\MeV}$ occurs
because branching ratio measurements provide bounds below this mass, and peak searches above this threshold, \textit{cf.}~discussion in Sec.~\ref{sec:exp}. Similarly, for $i=2$, the pion can 
decay into a muon instead of an electron, in the mass range, $m_{\tilde{\chi}^0_1} \le m_{\pi^{\pm}}
-m_{\mu}$. This time, the sensitivity range, shown in~\cref{fig:pion2}, comes from 
\texttt{SIN}~\cite{Daum:1987bg} in addition to \texttt{PIENU}~\cite{PIENU:2019usb} and \texttt{PIONEER}.
Finally, with $i=3$, the decays, $\tau^{\pm} \to \pi^{\pm} (\rho^\pm) + \tilde{\chi}^0_1$, can occur if 
kinematically allowed. While we could not find a direct search for such a process involving an HNL, 
$\lam'_{311}\not=0$ additionally allows the neutralino to decay into a tau neutrino and one of 
$\pi^0/\rho^0/\eta/\eta'/\omega$. Thus, displaced-vertex searches at \texttt{DUNE}~\cite{Ballett:2019bgd}, 
\texttt{FASER(2)}~\cite{Kling:2018wct} and \texttt{MoEDAL-MAPP2}~\cite{DeVries:2020jbs,Beltran:2023nli} 
can show sensitivity.\footnote{See also Ref.~\cite{Dey:2020juy} for a proposed search at Belle~II~\cite{Belle-II:2010dht,Belle-II:2018jsg} for a light bino with non-vanishing RPV couplings $\lambda'_{311}$ or $\lambda'_{312}$.\label{footenote:belleII_LQD}}
We present the corresponding combined projected sensitivity reach in Fig.~\ref{fig:pion3}. In all the plots, we also show the best existing constraints on the RPV couplings for different sfermion masses.
For $\lam'_{111}$, this comes from neutrinoless double beta decay searches~\cite{Hirsch:1998kc,Bolton:2021hje}, and is rather stringent compared to the other two couplings.
Nevertheless, we see that the reinterpreted bounds easily outperform these existing constraints in the major part of the phase space region.
We also note that in all of the above scenarios, the RPV couplings contribute to the invisible decay width of the pion, via the decay into a neutrino and the bino, given the long lifetime of the latter. However, the limits obtained this way are not competitive compared with the above ones.

\begin{figure}[!ht]
\begin{subfigure}{0.49\textwidth}
  \includegraphics[width=\linewidth]{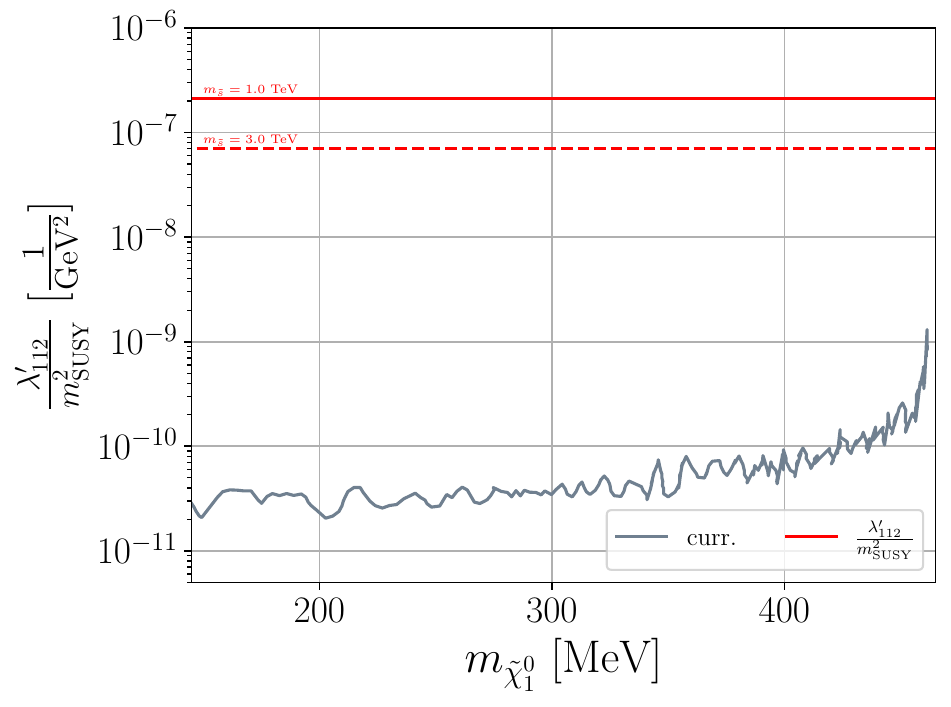}
  \caption{$\lambda'_{112}$.}
  \label{fig:kaon1}
  \end{subfigure}
  \begin{subfigure}{0.49\textwidth}
  \includegraphics[width=\linewidth]{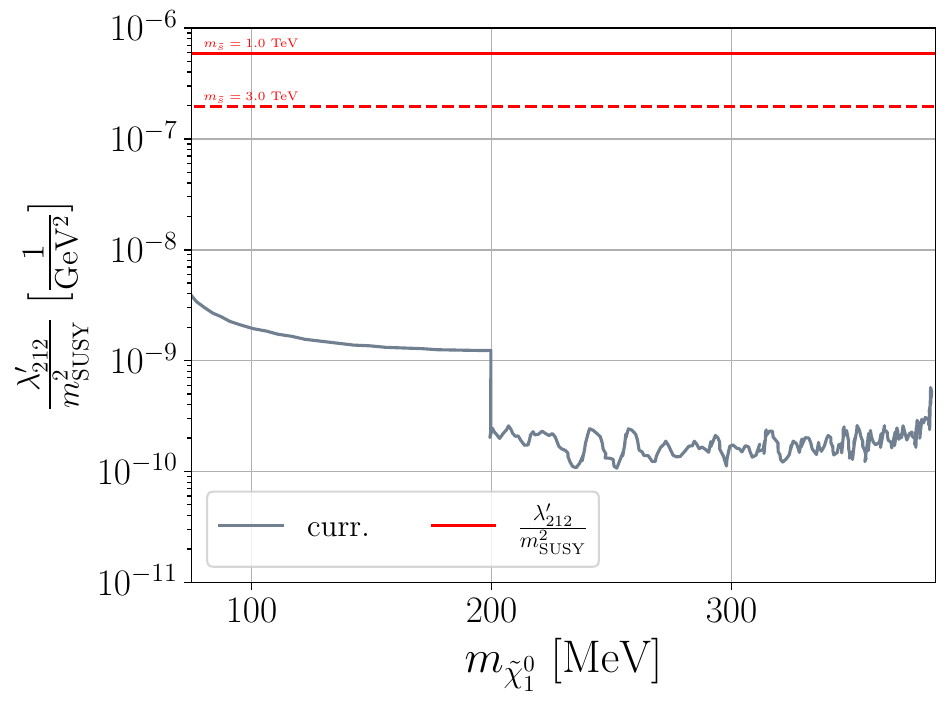}
  \caption{$\lambda'_{212}$.}
  \label{fig:kaon2}
  \end{subfigure}
\\
\centering
\begin{subfigure}{0.49\textwidth}
  \includegraphics[width=\linewidth]{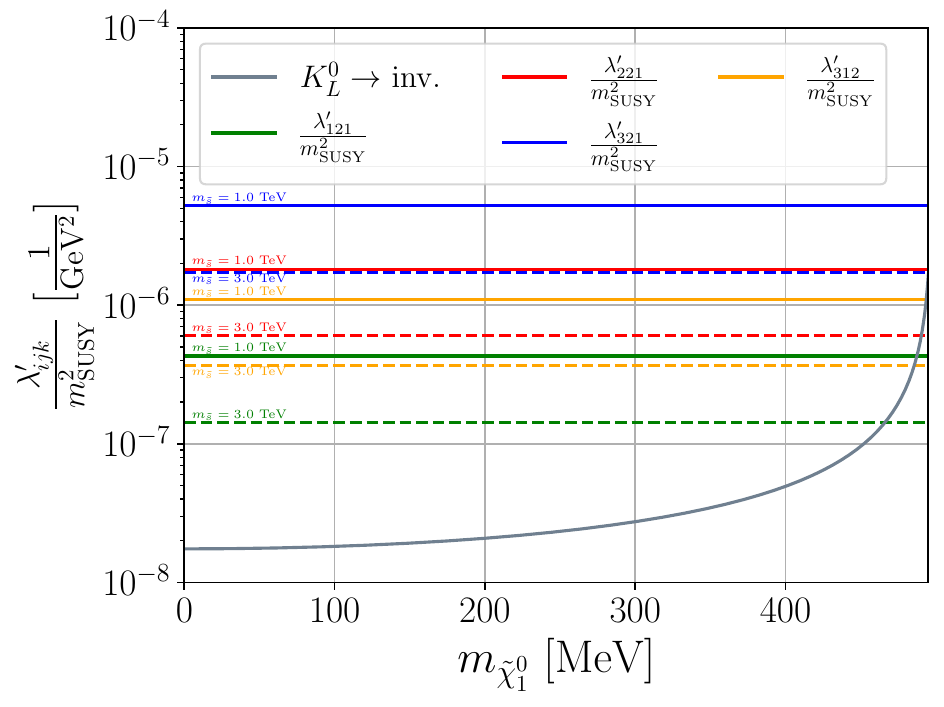}
  \caption{$\lambda'_{312}$ and $\lambda'_{i21}$.}
  \label{fig:kaon3}
  \end{subfigure}
\caption{As in~\cref{fig:pion_single} but for the $\bm{\lambda'_{K}}$ and $\bm{\lambda'_{D/K}}$ benchmarks of~\cref{tab:single_couplings_BMs}. The existing limits on the RPV couplings are shown in red, yellow, green, and blue.
}
\end{figure}

Next, we consider the kaon scenarios, involving couplings of the type $\lambda'_{i12}$. For 
$i\in\{1,2\}$, as for the pion case above, the decay mode into an electron or a muon along 
with the bino opens up for the kaon, below the relevant kinematic thresholds. The most 
constraining current limits come from \texttt{KEK}~\cite{Asano:1981he, Hayano:1982wu} and 
\texttt{NA62}~\cite{NA62:2020mcv, NA62:2021bji}, and are depicted in Fig.~\ref{fig:kaon1} 
(electron case), and Fig.~\ref{fig:kaon2} (muon case).
The degradation at $m_{\tilde{\chi}^0_1}\sim \SI{200}{\MeV}$ in Fig.~\ref{fig:kaon2} occurs 
because the \texttt{KEK}-limit only applies up to this mass; beyond it the \texttt{NA62} 
limit applies. Once again, both $L_iQ_1\bar D_2$ also contribute to the invisible decays of the kaon but the 
resulting limits are weaker than those shown. However, for $\lam'_{312}$ 
-- where we could not find an existing HNL direct search for the charged decay, $\tau^\pm \to 
K^\pm + \tilde{\chi}^0_1$ (see Footnote~\ref{footenote:belleII_LQD}), the invisible width of $K^0_L$ can indeed be used to derive limits. 
Since there is no direct bound on this width, we use the uncertainty on the total measured 
width of $K^0_L$~\cite{ParticleDataGroup:2022pth} to estimate the upper bound, $\text{BR}(K^0_L \to 
\text{invis.}) < 4.1 \times 10^{-3}$. This, then, can be used to constrain $\lambda'_{312}$ 
since it induces the decay of $K^0_L$ into a neutrino and a bino. Analogously, the invisible 
width also provides limits on couplings of the type $\lambda'_{i21}$. Here also, we could 
not find existing direct searches for the relevant charged decay of $D$ into a lepton and 
an HNL. All these constraints are displayed in Fig.~\ref{fig:kaon3}. Once again, we can see 
that the reinterpreted bounds exclude regions of phase space orders of magnitude beyond 
those ruled out by the existing constraints on the RPV operators (also shown in the plots).

\begin{figure}[!ht]
\begin{subfigure}{0.49\textwidth}
  \includegraphics[width=\linewidth]{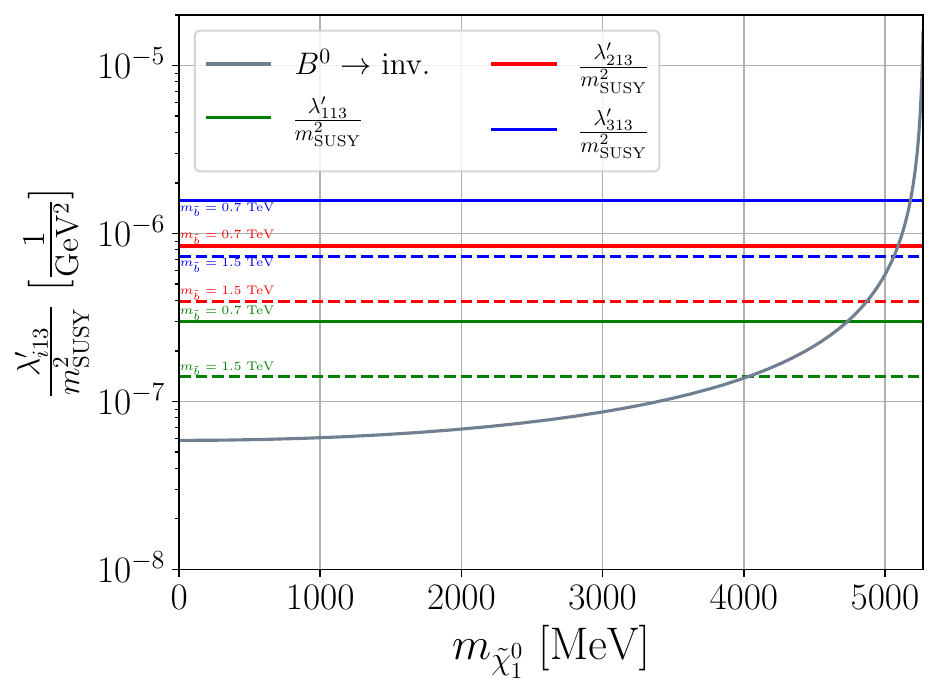}
  \caption{$\lambda'_{i13}$.}
  \label{fig:B1}
  \end{subfigure}
  \begin{subfigure}{0.49\textwidth}
  \includegraphics[width=\linewidth]{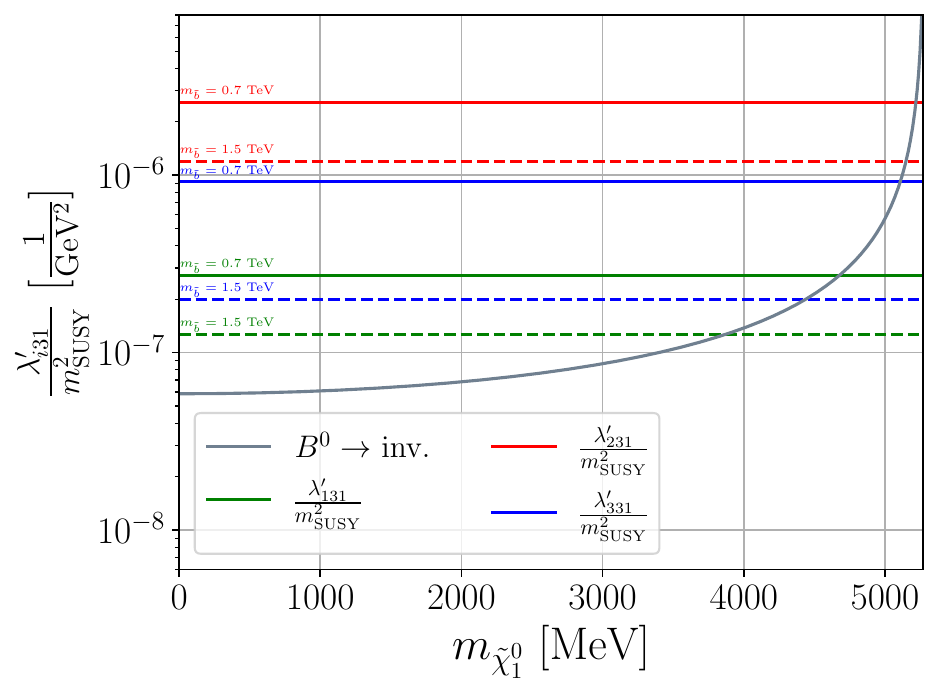}
  \caption{$\lambda'_{i31}$.
  }
  \label{fig:B2}
  \end{subfigure}
\caption{As in~\cref{fig:pion_single} but for the $\bm{\lambda'_{B1}}$ and $\bm{\lambda'_{B2}}$ benchmarks of~\cref{tab:single_couplings_BMs}. The existing limits on the RPV couplings are shown in red, green, and blue.
}
\label{fig:B}
\end{figure}

Similarly, for couplings of the type $\lambda'_{i13}$ and $\lambda'_{i31}$, we can use the 
invisible width of $B^0$. Here, direct measurements at \texttt{BaBar}~\cite{BaBar:2012yut} 
provide the stringent constraint, $\text{BR}(B^0 \to \text{invis.}) < 2.4\times 10^{-5}$. 
The resulting bounds are shown in Fig.~\ref{fig:B}.

\begin{figure}[!ht]
\begin{subfigure}{0.48\textwidth}
  \includegraphics[width=\linewidth]{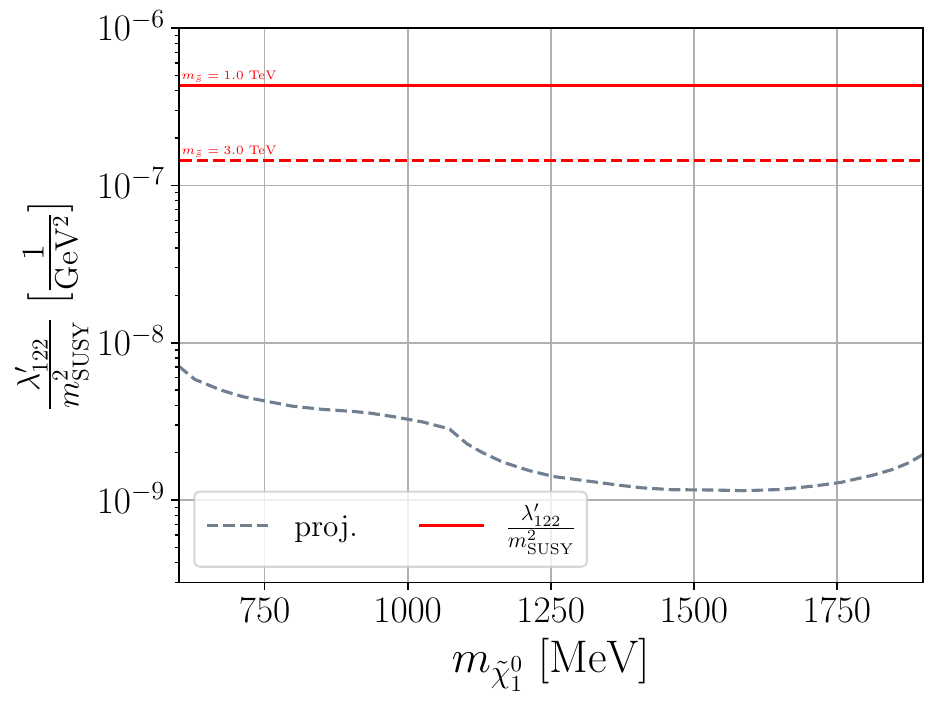}
  \caption{$\lambda'_{122}$.}
  \label{fig:Ds1}
  \end{subfigure}\hfill
\begin{subfigure}{0.48\textwidth}%
    \includegraphics[width=\linewidth]{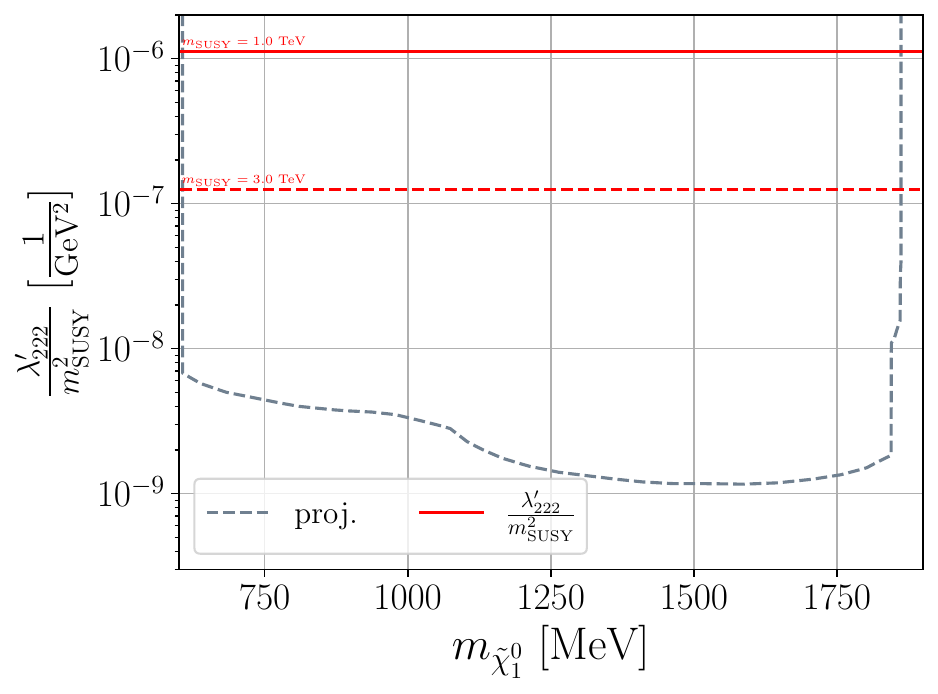}
\caption{$\lambda'_{222}$.}
  \label{fig:Ds2}
\end{subfigure}
\caption{As in~\cref{fig:pion_single} but for the $\bm{\lambda'_{D_s}}$ benchmarks of~\cref{tab:single_couplings_BMs}.
}
\label{fig:Ds}
\end{figure}

For the couplings $\lam'_{a22}$, with $a\in\{1,2\}$, we use potential future displaced-vertex 
searches from \texttt{DUNE}~\cite{Ballett:2019bgd}, \texttt{FASER2}~\cite{Kling:2018wct}, and 
\texttt{MoEDAL-MAPP2}~\cite{DeVries:2020jbs,Beltran:2023nli} since we could not find a direct 
search for the decay of $D^\pm_s$ into an HNL. In these scenarios, the neutralino decays via the 
RPV operator into a neutrino and one of $\phi/\eta/\eta'$. This implies the 
mass range, $\SI{548}{\MeV}\lesssim m_{\tilde{\chi}^0_1} < m_{D_s}-m_{e_a}$. Thus, the corresponding 
scenario with $\lam'_{322}\not=0$ involving a tau lepton is not possible. We show the resulting 
sensitivity limits in Fig.~\ref{fig:Ds}.

For the remaining $LQ\Bar{D}$ operators, there are either no relevant processes providing constraints through meson or lepton decays ($\lambda'_{322}, \lambda'_{i33}$), or the obtained limits are not competitive with the existing bounds ($\lambda'_{i23}, \lambda'_{i32}$).

\subsubsection{$LL\Bar{E}$ scenarios} 
\begin{figure}[!ht]
\centering	    
\includegraphics[width=0.6\textwidth]{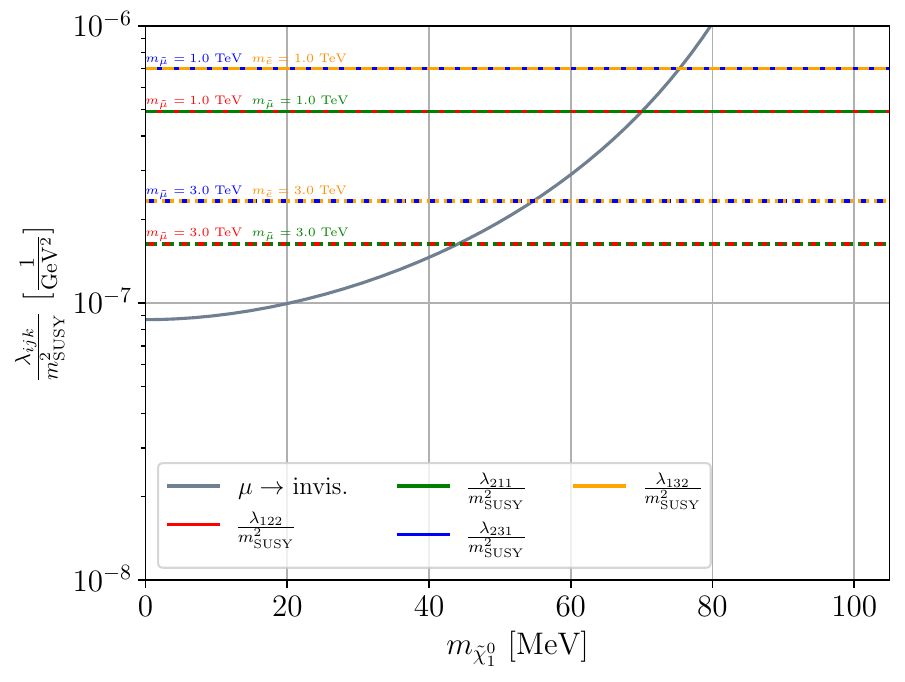}
\caption{As in~\cref{fig:pion_single} but for the $\bm{\lambda'_{\mu}}$ and $\bm{\lambda'_{\tau/\mu}}$ benchmarks of~\cref{tab:single_couplings_BMs}. The existing limits on the RPV couplings are shown in red, yellow, green, and blue.}
\label{fig:LLE}
\end{figure}

For the $\bm{\lam_{\mu}}, \bm{\lam_{\tau}}$ and $\bm{\lam_{\tau/\mu}}$ scenarios 
of~\cref{tab:single_couplings_BMs}, the corresponding operators 
contribute to the leptonic decays of the muon and the tau. Since we could not find direct 
measurements for the process $\mu^\pm\to e^\pm + \text{invis.}$ (and analogously for $\tau\to 
e$ and $\tau \to \mu$), where the invisible final state may be massive, we use the uncertainty
on the muon decay width~\cite{ParticleDataGroup:2022pth} to obtain an estimated bound, $\text{BR}(\mu^\pm
\to e^\pm +\text{invis.}) < 1.0\times 10^{-6}$. This leads to
constraints on the $\bm{\lam_{\mu}}$ and $\bm{\lam_{\tau/\mu}}$ scenarios, as shown in~\cref{fig:LLE}.
The analogous procedure with $\tau$ leads to limits weaker than the existing constraints and 
are hence not presented here.

\subsection{Two-coupling scenarios}

Next, we consider scenarios where two RPV operators are simultaneously switched on; one corresponds to the production of the bino LSP, while the other leads to the decay. We 
can classify these scenarios based on the parent particle producing the neutralino; this then fixes 
the production RPV coupling. The relevant possibilities for the parent particles at beam-dump searches
and colliders are: pions, kaons, $D$ mesons, $D_s$ mesons, $B$ mesons, $B_c$ 
mesons, and $\tau$ leptons.\footnote{$\tau$ leptons are dominantly produced in the decay of the 
$D_s$ mesons at these experiments.} The corresponding production RPV couplings (and processes) 
can be read off from~\cref{tab:single_couplings_BMs}, \textit{e.g.}, $\lambda'_{a11}$ for the 
pion, etc.

The produced neutralino, owing to its long lifetime, then travels a certain macroscopic distance and 
decays via the other RPV coupling. Thus, displaced-vertex searches for HNLs are sensitive to such scenarios, if the final states match the decay products of 
the bino. We list, in~\cref{tab:decay_modes_HNL}, all relevant final states of such 
searches, along with the relevant kinematic thresholds at which the HNL can decay 
into them. Consulting the discussion in Sec.~\ref{sec:model}, we see that each of these final states 
can also arise from bino decays.

\begin{table}[!ht]
   \centering
\begin{tabular}{|lr|lr|}
\hline
Channel & Threshold & Channel & Threshold  \\
\hline
$\nu e^+ e^-$ & \SI{1.02}{\MeV} &$\mu^{\mp} K^{ \pm}$ & \SI{599}{\MeV} \\
$\nu e^{ \pm} \mu^{\mp}$ & \SI{105}{\MeV} & $\nu \rho^0$ & \SI{776}{\MeV} \\
$\nu \pi^0$ & \SI{135}{\MeV} & $e^{\mp} \rho^{ \pm}$ & \SI{776}{\MeV} \\
$e^{\mp} \pi^{ \pm}$ & \SI{140}{\MeV} & $\nu \omega$ & \SI{783}{\MeV} \\
$\nu \mu^{+} \mu^{-}$ & \SI{210}{\MeV} & $\mu^{\mp} \rho^{ \pm}$ & \SI{882}{\MeV} \\
$\mu^{\mp} \pi^{ \pm}$ & \SI{245}{\MeV} & $\nu \eta^{\prime}$ & \SI{958}{\MeV} \\
$e^{\mp} K^{ \pm}$ &  \SI{494}{\MeV} & $\nu \phi$ & \SI{1019}{\MeV} \\
$\nu \eta$ & \SI{548}{\MeV} &&\\
\hline
\end{tabular}
 \caption{Relevant final states from HNL (and bino) decay sorted by threshold mass. The active neutrino is considered massless.}
 \label{tab:decay_modes_HNL}
 \end{table}
In~\cref{tab:two_coupling_BMs}, we list, for each production category, the relevant RPV operator(s) leading to the final states of~\cref{tab:decay_modes_HNL}.
$\bm{\times}$ indicates that the given final state can not arise for the considered production mode, owing to kinematics.
This table can be used to identify all relevant two-coupling RPV scenarios that can be constrained by existing HNL DV searches. Note that we have not included bino production modes corresponding to the $LL\Bar{E}$ decays of the $\tau$ leptons in the table since these lead to weak limits, as in the one-coupling scenarios. We now discuss numerical results for a representative subset of the possibilities in the table. 

\subsubsection{Pion scenarios} 
For the $\pi^\pm$ category, we consider the two benchmark scenarios listed in~\cref{tab:pion_BMs}. 
The relevant mass range is identified by requiring that the production of the neutralino, and its 
subsequent decay (\textit{cf.}~\cref{tab:two_coupling_BMs}), should be both kinematically accessible.

We use this opportunity to explain one more subtlety of our recasting procedure for DV searches. In the HNL model, the relevant pion decay process occurs via the neutrino-mixings, 
\textit{e.g.}, $\pi^\pm \to e^\pm + N$ via $U_e$. However, a non-zero $U_e$ also induces 
the production of the HNL via decays of other particles, \textit{e.g.}, $K^\pm \to e^\pm 
+ N$ and the three-body decay, $K^\pm \to e^\pm + \pi^0 + N$. On the other hand, in our RPV 
benchmarks, this is not the case since $\lambda'_{i11}$ only couples to the pion. This affects
the kinematics of the HNL relative to the bino, and can be an issue for our simple scaling 
procedure, \textit{cf.} the discussion in Sec.~\ref{sec:exp}. For the pion benchmarks 
of~\cref{tab:pion_BMs}, this is not a problem since the kaon modes are sub-dominant $\left(
\mathcal{O}(1\%)\right)$ for the mass range identified above: $\SI{1}{\MeV} \lesssim m_N \lesssim 
\SI{139}{\MeV}$; see, for instance, Ref.~\cite{Berryman:2019dme} for a plot with the relevant 
branching ratios in the HNL scenario. However, in later benchmarks, we deal with this issue 
-- if it arises -- by restricting the mass range of the benchmark to ensure that the types of 
contributing parents are the same in both models (neglecting sub-dominant contributions up to 
$\mathcal{O}(10\%)$), so that the assumption of the same kinematics taken in the DV quick 
recasting method still holds. 

\begin{table}[H]
\centering
\begin{adjustbox}{width=0.75\columnwidth}
\rotatebox{90}{
\begin{tabular}{|c|c|c|c|c|c|c|c|c|c|c|c|}
\hline
\bf{Category} & \diagbox[width=5cm]{$\bm{\tilde{\chi}^0_1}$ \bf{Production}}{$\bm{\tilde{\chi}^0_1}$ \bf{Decay}} & $\bm{\nu ee}$ & $\bm{\nu e\mu}$ & $\bm{\nu \mu \mu}$ & $\bm{e\{\pi/\rho\}}$ & $\bm{\mu\{\pi/\rho\}}$ & $\bm{eK}$ & $\bm{\mu K}$ & $\bm{\nu\{\pi/\rho/\omega\}}$ & $\bm{\nu\{\eta/\eta'/\phi\}}$\\[0.15 cm]
\hline
\multirow{4}{*}{$\pi$} & $\lambda'_{111}: \pi^\pm \to e^\pm + \tilde{\chi}^0_1$ & \multirowcell{2}{$\lambda_{121}; \lambda_{131}$} & \multirowcell{2}{$\lambda_{121}; \lambda_{122}; \lambda_{321}; \lambda_{312}$} & \multirow{2}{*}{$\bm{\times}$} & \multirow{2}{*}{$\bm{\times}$} & \multirow{2}{*}{$\bm{\times}$} & \multirow{2}{*}{$\bm{\times}$} & \multirow{2}{*}{$\bm{\times}$} & \multirow{2}{*}{$\bm{\times}$} & \multirow{2}{*}{$\bm{\times}$}\\[0.15 cm]
& $(m_{\tilde{\chi}^0_1} \lesssim \SI{139}{\MeV})$ & & & & & & & & &\\[0.15 cm]
& $\lambda'_{211}: \pi^\pm \to \mu^\pm + \tilde{\chi}^0_1$ & \multirowcell{2}{$\lambda_{121}; \lambda_{131}$} & \multirowcell{2}{$\bm{\times}$} & \multirow{2}{*}{$\bm{\times}$} & \multirow{2}{*}{$\bm{\times}$} & \multirow{2}{*}{$\bm{\times}$} & \multirow{2}{*}{$\bm{\times}$} & \multirow{2}{*}{$\bm{\times}$} & \multirow{2}{*}{$\bm{\times}$} & \multirow{2}{*}{$\bm{\times}$}\\[0.15 cm]
& $(m_{\tilde{\chi}^0_1} \lesssim \SI{34}{\MeV})$ & & & & & & & & &\\[0.15 cm]
\hline
\multirow{4}{*}{$K$} & $\lambda'_{112}: K^\pm \to e^\pm + \tilde{\chi}^0_1$ & \multirowcell{2}{$\lambda_{121}; \lambda_{131}$} & \multirowcell{2}{$\lambda_{121}; \lambda_{122}; \lambda_{321}; \lambda_{312}$} & \multirow{2}{*}{$\lambda_{122}; \lambda_{322}$} & \multirow{2}{*}{$\lambda'_{111}$} & \multirow{2}{*}{$\lambda'_{211}$} & \multirow{2}{*}{$\bm{\times}$} & \multirow{2}{*}{$\bm{\times}$} & \multirow{2}{*}{$\lambda'_{i11}$} & \multirow{2}{*}{$\bm{\times}$}\\[0.15 cm]
& $(m_{\tilde{\chi}^0_1} \lesssim \SI{493}{\MeV})$ & & & & & & & & &\\[0.15 cm]
& $\lambda'_{212}: K^\pm \to \mu^\pm + \tilde{\chi}^0_1$ & \multirowcell{2}{$\lambda_{121}; \lambda_{131}$} & \multirowcell{2}{$\lambda_{121}; \lambda_{122}; \lambda_{321}; \lambda_{312}$} & \multirow{2}{*}{$\lambda_{122}; \lambda_{322}$} & \multirow{2}{*}{$\lambda'_{111}$} & \multirow{2}{*}{$\lambda'_{211}$} & \multirow{2}{*}{$\bm{\times}$} & \multirow{2}{*}{$\bm{\times}$} & \multirow{2}{*}{$\lambda'_{i11}$} & \multirow{2}{*}{$\bm{\times}$}\\[0.15 cm]
& $(m_{\tilde{\chi}^0_1} \lesssim \SI{388}{\MeV})$ & & & & & & & & &\\[0.15 cm]
\hline
\multirow{2}{*}{$D$} & $\lambda'_{a21}: D^\pm \to e_a^\pm + \tilde{\chi}^0_1$ & \multirowcell{2}{$\lambda_{121}; \lambda_{131}$} & \multirowcell{2}{$\lambda_{121}; \lambda_{122}; \lambda_{321}; \lambda_{312}$} & \multirow{2}{*}{$\lambda_{122}; \lambda_{322}$} & \multirow{2}{*}{$\lambda'_{111}$} & \multirow{2}{*}{$\lambda'_{211}$} & \multirow{2}{*}{$\lambda'_{112}$} & \multirow{2}{*}{$\lambda'_{212}$} & \multirow{2}{*}{$\lambda'_{i11}$} & \multirow{2}{*}{$\lambda'_{i11}; \lambda'_{i22}$}\\[0.15 cm]
& $(m_{\tilde{\chi}^0_1} \le m_{D}-m_{e_a})$ & & & & & & & & &\\[0.15 cm]
\hline
\multirow{2}{*}{$D_s$} & $\lambda'_{a22}: D_s^\pm \to e_a^\pm + \tilde{\chi}^0_1$ & \multirowcell{2}{$\lambda_{121}; \lambda_{131}$} & \multirowcell{2}{$\lambda_{121}; \lambda_{122}; \lambda_{321}; \lambda_{312}$} & \multirow{2}{*}{$\lambda_{122}; \lambda_{322}$} & \multirow{2}{*}{$\lambda'_{111}$} & \multirow{2}{*}{$\lambda'_{211}$} & \multirow{2}{*}{$\lambda'_{112}$} & \multirow{2}{*}{$\lambda'_{212}$} & \multirow{2}{*}{$\lambda'_{i11}$} & \multirow{2}{*}{$\lambda'_{i11}; \lambda'_{322}$}\\[0.15 cm]
& $(m_{\tilde{\chi}^0_1} \le m_{D_s}-m_{e_a})$ & & & & & & & & &\\[0.15 cm]
\hline
\multirow{2}{*}{$B$} & $\lambda'_{i13}: B^\pm(B^0) \to e_i^\pm(\nu_i) + \tilde{\chi}^0_1$ & \multirowcell{2}{$\lambda_{121}; \lambda_{131}$} & \multirowcell{2}{$\lambda_{121}; \lambda_{122}; \lambda_{321}; \lambda_{312}$} & \multirow{2}{*}{$\lambda_{122}; \lambda_{322}$} & \multirow{2}{*}{$\lambda'_{111}$} & \multirow{2}{*}{$\lambda'_{211}$} & \multirow{2}{*}{$\lambda'_{112}$} & \multirow{2}{*}{$\lambda'_{212}$} & \multirow{2}{*}{$\lambda'_{i11}$} & \multirow{2}{*}{$\lambda'_{i11}; \lambda'_{i22}$}\\[0.15 cm]
& $(m_{\tilde{\chi}^0_1} \le m_{B}-m_{e_i})$ & & & & & & & & &\\[0.15 cm]
\hline
\multirow{2}{*}{$B_c$} & $\lambda'_{i23}: B_c^\pm \to e_i^\pm + \tilde{\chi}^0_1$ & \multirowcell{2}{$\lambda_{121}; \lambda_{131}$} & \multirowcell{2}{$\lambda_{121}; \lambda_{122}; \lambda_{321}; \lambda_{312}$} & \multirow{2}{*}{$\lambda_{122}; \lambda_{322}$} & \multirow{2}{*}{$\lambda'_{111}$} & \multirow{2}{*}{$\lambda'_{211}$} & \multirow{2}{*}{$\lambda'_{112}$} & \multirow{2}{*}{$\lambda'_{212}$} & \multirow{2}{*}{$\lambda'_{i11}$} & \multirow{2}{*}{$\lambda'_{i11}; \lambda'_{i22}$}\\[0.15 cm]
& $(m_{\tilde{\chi}^0_1} \le m_{B_c}-m_{e_i})$ & & & & & & & & &\\[0.15 cm]
\hline
\multirow{4}{*}{$\tau$} & $\lambda'_{311}: \tau^\pm \to \{\pi^\pm/\rho^\pm\} + \tilde{\chi}^0_1$ & \multirowcell{2}{$\lambda_{121}; \lambda_{131}$} & \multirowcell{2}{$\lambda_{121}; \lambda_{122}; \lambda_{321}; \lambda_{312}$} & \multirow{2}{*}{$\lambda_{122}; \lambda_{322}$} & \multirow{2}{*}{$\lambda'_{111}$} & \multirow{2}{*}{$\lambda'_{211}$} & \multirow{2}{*}{$\lambda'_{112}$} & \multirow{2}{*}{$\lambda'_{212}$} & \multirow{2}{*}{$\lambda'_{a11}$} & \multirow{2}{*}{$\lambda'_{a11}; \lambda'_{i22}$}\\[0.15 cm]
& $(m_{\tilde{\chi}^0_1} \lesssim \SI{1637}{\MeV})$ & & & & & & & & &\\[0.15 cm]
& $\lambda'_{312}: \tau^\pm \to K^\pm + \tilde{\chi}^0_1$ & \multirowcell{2}{$\lambda_{121}; \lambda_{131}$} & \multirowcell{2}{$\lambda_{121}; \lambda_{122}; \lambda_{321}; \lambda_{312}$} & \multirow{2}{*}{$\lambda_{122}; \lambda_{322}$} & \multirow{2}{*}{$\lambda'_{111}$} & \multirow{2}{*}{$\lambda'_{211}$} & \multirow{2}{*}{$\lambda'_{112}$} & \multirow{2}{*}{$\lambda'_{212}$} & \multirow{2}{*}{$\lambda'_{i11}$} & \multirow{2}{*}{$\lambda'_{i11}; \lambda'_{i22}$}\\[0.15 cm]
& $(m_{\tilde{\chi}^0_1} \lesssim \SI{1283}{\MeV})$ & & & & & & & & &\\[0.15 cm]
\hline
\end{tabular}}
\end{adjustbox}
\caption{Relevant two-coupling RPV scenarios probed by HNL DV searches. Column one categorizes the parent meson/lepton; column two shows the 
corresponding RPV coupling, production process, and bino mass range; the remaining columns list the relevant final states of~\cref{tab:decay_modes_HNL} arising from bino decays and the corresponding decay RPV coupling(s) for each production category. $\bm{\times}$ indicates 
that the decay is kinematically disallowed.}
\label{tab:two_coupling_BMs}
\end{table}

\begin{table}
\centering
\begin{tabular}{|c|c|c|c|}
\hline
\bf{Label} & \bf{Production} & \bf{Decay} & $\bm{m_{\tilde{\chi}^0_1}}$\\
\hline
$\bm{\pi_1}$ & $\lambda'_{111}$ & $\lambda_{321}$ & $\SI{106}{\MeV} - \SI{139}{\MeV}$\\
$\bm{\pi_2}$ & $\lambda'_{211}$ & $\lambda_{121}$ & $\SI{1}{\MeV}-\SI{34}{\MeV}$\\
\hline
\end{tabular}
\caption{Details of the two-coupling RPV benchmark scenarios we study, corresponding to bino 
production from pions. The decay coupling in the third column leads to the final state that 
can be read off from~\cref{tab:two_coupling_BMs}. See the main text for details on how the 
mass range is determined.}
\label{tab:pion_BMs}
\end{table}

\begin{figure}[!ht]
\begin{subfigure}{0.49\textwidth}
  \includegraphics[width=\linewidth]{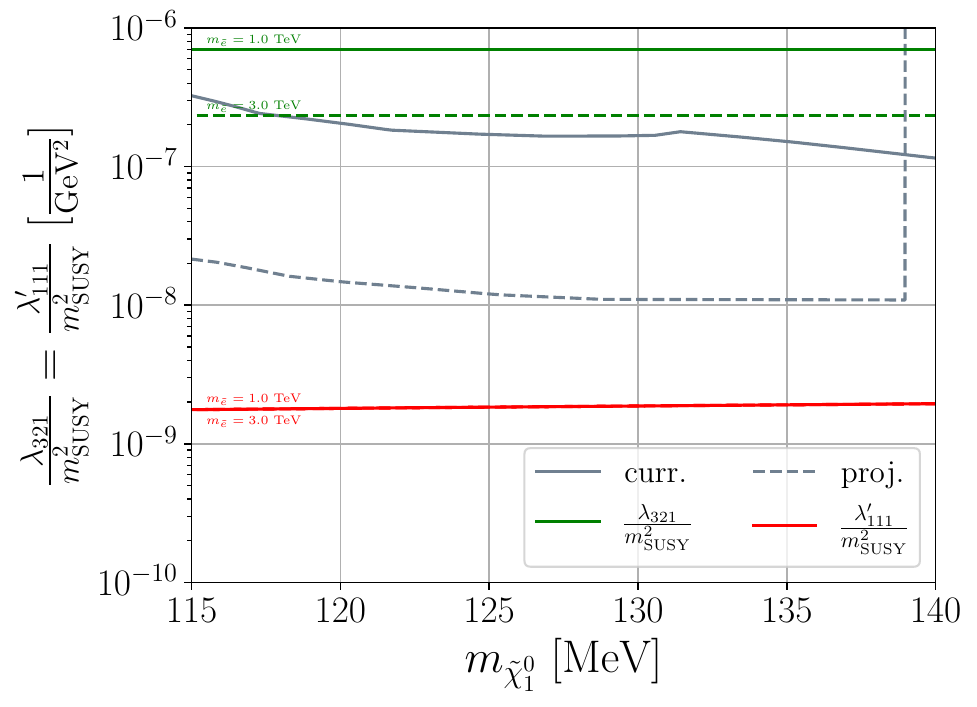}
  \caption{Limits in the RPV coupling vs.~bino mass plane for the benchmark $\bm{\pi_1}$ of~\cref{tab:pion_BMs}.}
  \label{fig:pion_two_1}
  \end{subfigure}\hfill
\begin{subfigure}{0.49\textwidth}%
    \includegraphics[width=\linewidth]{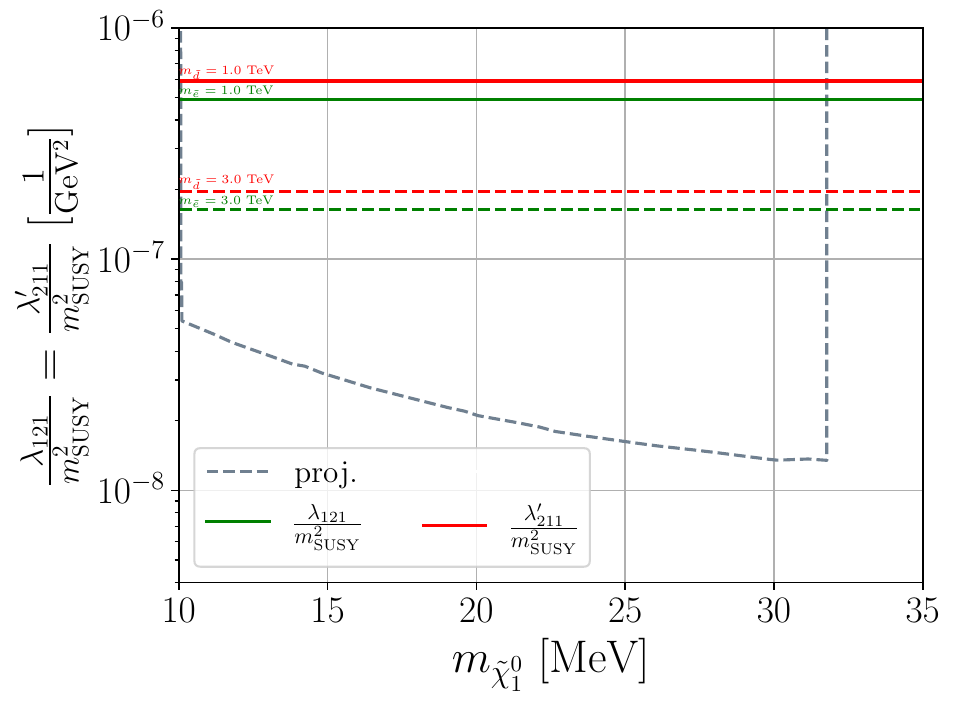}
  \caption{Limits in the RPV coupling vs.~bino mass plane for the benchmark $\bm{\pi_2}$ of~\cref{tab:pion_BMs}.}
  \label{fig:pion_two_2}
\end{subfigure}
\\
\centering
\begin{subfigure}{0.49\textwidth}
  \includegraphics[width=\linewidth]{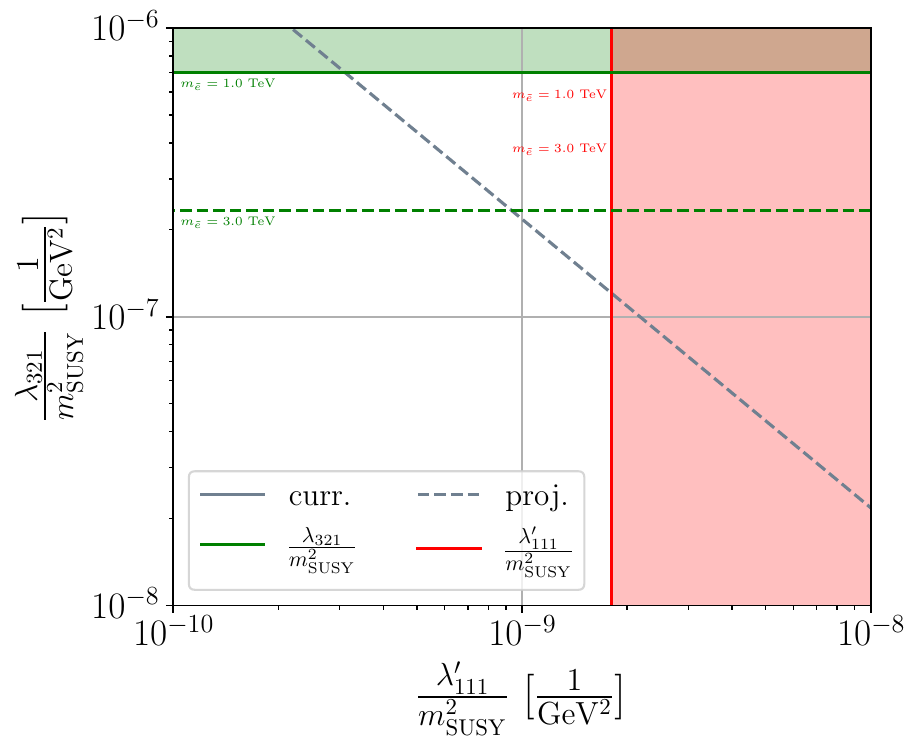}
  \caption{Limits in the production coupling vs.~decay coupling plane for the benchmark $\bm{\pi_1}$ of~\cref{tab:pion_BMs}, with neutralino mass fixed at \SI{120}{\MeV}.}
  \label{fig:pion_two_1a}
  \end{subfigure}
\caption{Current exclusion (solid lines) and projected sensitivity (dashed lines) limits corresponding to the two-coupling RPV scenarios with binos produced from pions; reinterpreted from HNL searches. The existing limits on the RPV couplings are shown in red, and green.}
\end{figure}

We show the reinterpreted limits for the benchmarks $\bm{\pi_1}$ and $\bm{\pi_2}$ in the RPV 
coupling vs.~mass plane in Fig.~\ref{fig:pion_two_1} and Fig.~\ref{fig:pion_two_2}, 
respectively. For two-dimensional visualization, we have set the production and decay 
couplings to be equal. The exclusion limits come from a corresponding HNL search at 
\texttt{Super-Kamiokande}~\cite{Coloma:2019htx}, as well as projections of sensitivity 
at \texttt{DUNE}~\cite{Ballett:2019bgd}. The former only constrains the mass range $m_N \gsim 
\SI{50}{\MeV}$ and, hence, does not have sensitivity to $\bm{\pi_2}$. We also show the 
existing bounds on the RPV couplings, taken from Ref.~\cite{Dercks:2017lfq}. There are no 
existing product bounds on the pairs considered \cite{Allanach:1999ic, Barbier:2004ez}. From 
Fig.~\ref{fig:pion_two_1}, it appears that the existing bound on $\lam'_{111}$ outperforms 
the reinterpreted bound. However, this is an artefact of the choice to set the production and decay couplings equal. We show the same exclusion limits again in 
Fig.~\ref{fig:pion_two_1a} -- this time in the coupling vs.~coupling plane for a fixed 
neutralino mass, $m_{\tilde{\chi}_1^0}$ = \SI{120}{\MeV}; one can see that the reinterpreted 
sensitivity projection can probe a small region of the phase space still allowed by the current 
limits.

\subsubsection{Kaon scenarios}
Next, we study benchmarks corresponding to bino production via $K^\pm$ decays; the 
details are summarized in~\cref{tab:kaon_BMs}. Note, that for some benchmarks, 
(\textit{e.g.}, $\bm{K_3}$), the lower end of the mass range lies significantly 
above the kinematic threshold requirement of bino decay. This is, as discussed 
above, to ensure that kaons are the only parents in the HNL model, as they are in 
the RPV model.

\begin{table}[h]
\centering
\begin{tabular}{|c|c|c|c|}
\hline
\bf{Label} & \bf{Production} & \bf{Decay} & $\bm{m_{\tilde{\chi}^0_1}}$\\
\hline
$\bm{K_1}$ & $\lam'_{112}$ & $\lam'_{111}$ & $\SI{140}{\MeV} - \SI{493}{\MeV}$\\
$\bm{K_2}$ & $\lam'_{112}$ & $\lam'_{311}$ & $\SI{140}{\MeV}-\SI{493}{\MeV}$\\
$\bm{K_3}$ & $\lam'_{112}$ & $\lam_{321}$ & $\SI{140}{\MeV} - \SI{493}{\MeV}$\\
$\bm{K_4}$ & $\lam'_{212}$ & $\lam'_{211}$ & $\SI{140}{\MeV}-\SI{388}{\MeV}$\\
$\bm{K_5}$ & $\lam'_{212}$ & $\lam_{131}$ & $\SI{35}{\MeV} - \SI{388}{\MeV}$\\
\hline
\end{tabular}
\caption{As in~\cref{tab:pion_BMs} but for bino production from kaons.}
\label{tab:kaon_BMs}
\end{table}

\begin{figure}[!h]
\begin{subfigure}{0.49\textwidth}
  \includegraphics[width=\linewidth]{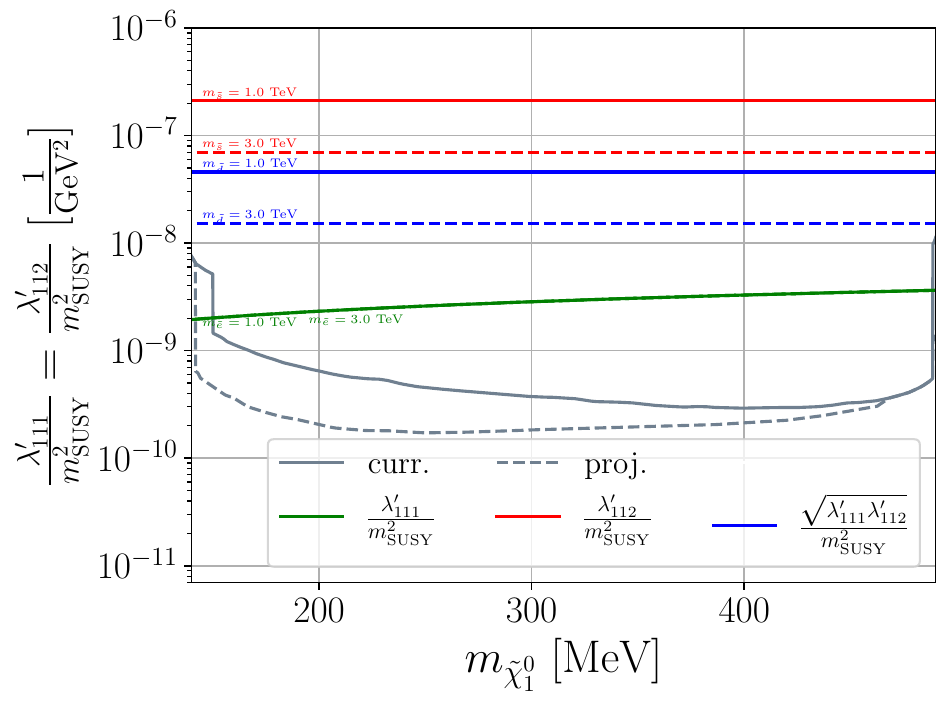}
  \caption{Benchmark $\bm{K_1}$ from~\cref{tab:kaon_BMs}.}
  \label{fig:kaon_two_1}
  \end{subfigure}\hfill
\begin{subfigure}{0.49\textwidth}%
    \includegraphics[width=\linewidth]{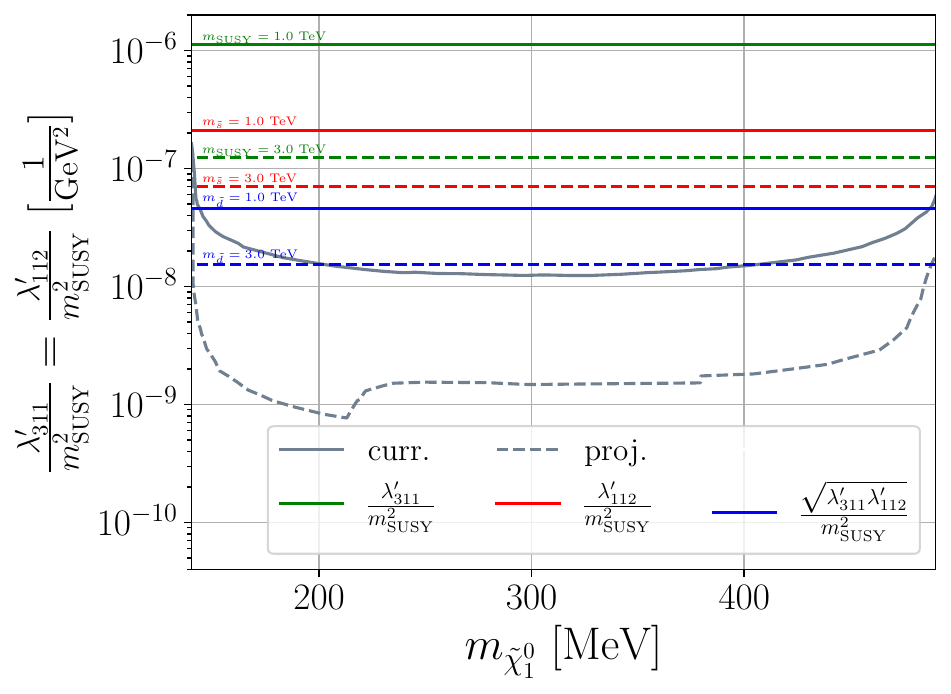}
  \caption{Benchmark $\bm{K_2}$ from~\cref{tab:kaon_BMs}.}
  \label{fig:kaon_two_2}
\end{subfigure}
\\
\begin{subfigure}{0.49\textwidth}
  \includegraphics[width=\linewidth]{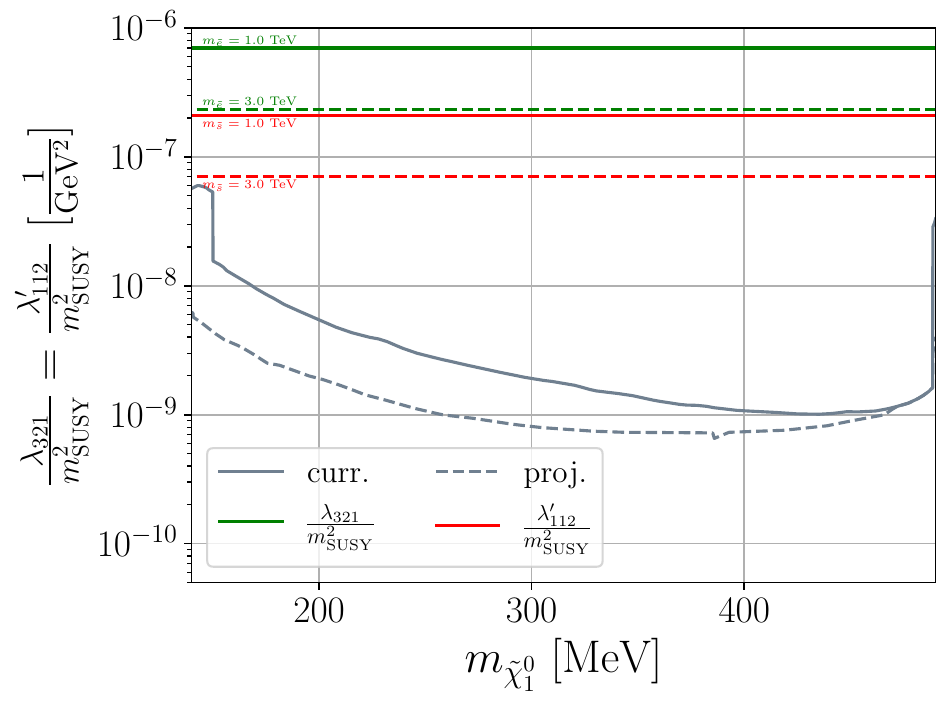}
  \caption{Benchmark $\bm{K_3}$ from~\cref{tab:kaon_BMs}.}
  \label{fig:kaon_two_3}
  \end{subfigure}\hfill
\begin{subfigure}{0.49\textwidth}%
    \includegraphics[width=\linewidth]{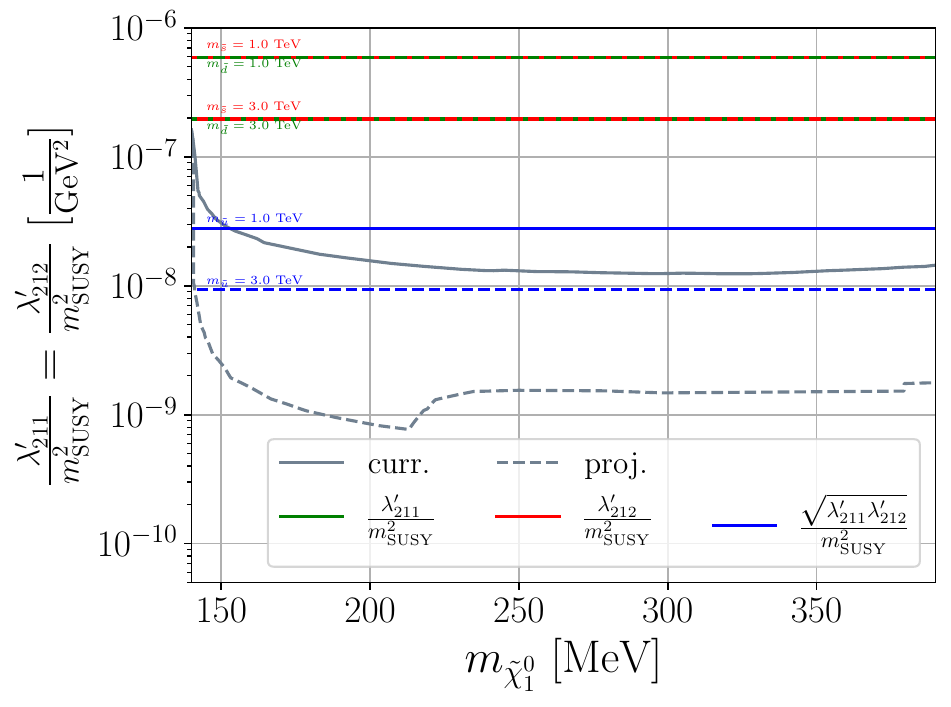}
  \caption{Benchmark $\bm{K_4}$ from~\cref{tab:kaon_BMs}.}
  \label{fig:kaon_two_4}
\end{subfigure}
\\
\centering
\begin{subfigure}{0.49\textwidth}
  \includegraphics[width=\linewidth]{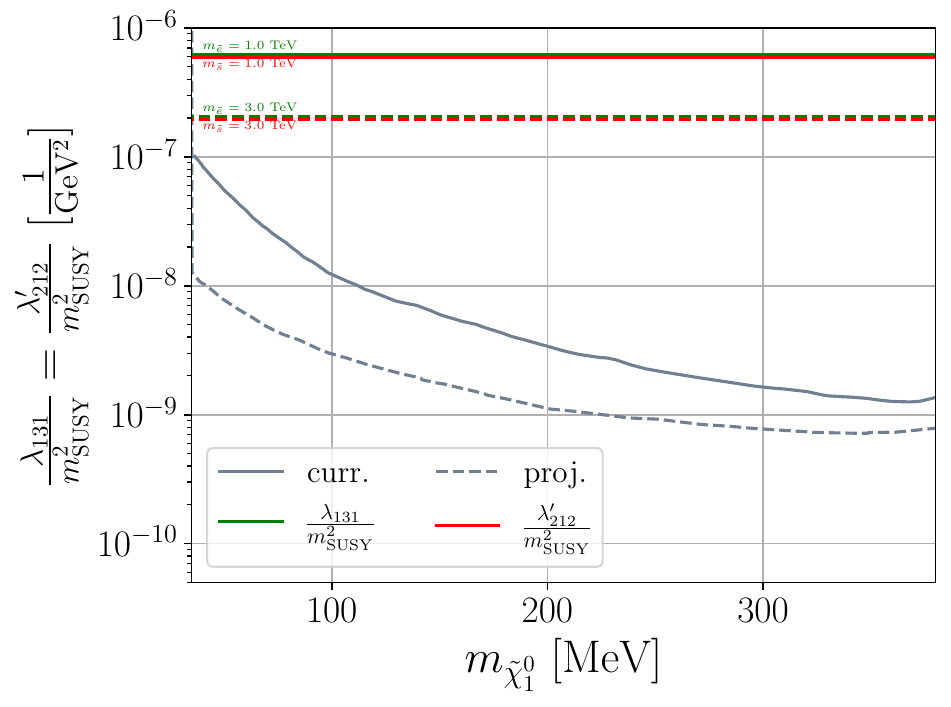}
  \caption{Benchmark $\bm{K_5}$ from~\cref{tab:kaon_BMs}.}
  \label{fig:kaon_two_5}
  \end{subfigure}
\caption{Current exclusion (solid lines) and projected sensitivity (dashed lines) 
limits corresponding to the two-coupling RPV scenarios with binos produced from 
kaons in the RPV coupling vs.~bino mass plane; reinterpreted from HNL searches. The existing limits on the RPV couplings are shown in red and green (single bounds), and blue (product bound).
}
\label{fig:kaon_two}
\end{figure}

The sensitivity limits for the kaon benchmarks are shown in Fig.~\ref{fig:kaon_two}. The 
single and -- wherever relevant -- product bounds (taken from Ref.~\cite{Allanach:1999ic, 
Barbier:2004ez}) on RPV couplings are also shown. Current exclusion limits are obtained 
by combining the results from existing HNL searches at \texttt{T2K}~\cite{T2K:2019jwa}, 
\texttt{Super-Kamiokande}~\cite{Coloma:2019htx}, \texttt{NuTeV}~\cite{NuTeV:1999kej}, 
and \texttt{MicroBooNE}~\cite{Kelly:2021xbv, MicroBooNE:2022ctm}, while the projections 
are all from \texttt{DUNE}~\cite{Ballett:2019bgd}.
In particular, for the benchmark $\bm{K_1}$, Ref.~\cite{Candia:2021bsl} has studied the sensitivity of \texttt{Super-Kamiokande} to the light binos in the RPV-SUSY, using existing data from the experiment.
Their results are found to be comparable with ours.
Once again, we see that the reinterpreted limits exclude (or are projected to probe) large swathes of 
parameter space allowed by the current bounds. The sharp reduction in sensitivity 
in~\cref{fig:kaon_two_5} below $m_{\tilde{\chi}^0_1} \approx \SI{150}{\MeV}$ 
arises because the most constraining current limit comes from \texttt{T2K} and only probes regions corresponding to $m_N \gsim \SI{150}{\MeV}$; below this the low-mass searches from \texttt{Super-K} and \texttt{MicroBooNE} provide exclusion.

\subsubsection{$D$, ${D}_s$, and $\tau$ scenarios} 
\begin{figure}[h]
\begin{subfigure}{0.49\textwidth}
  \includegraphics[width=\linewidth]{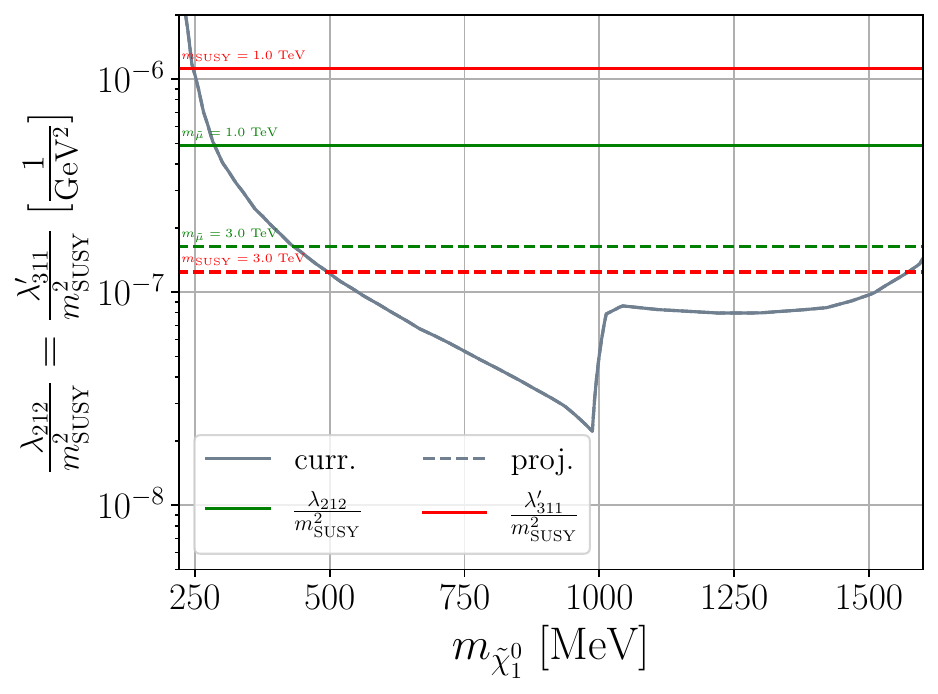}
  \caption{Benchmark $\bm{\tau_1}$ from~\cref{tab:d_BMs}.}
  \label{fig:tau_two_1}
  \end{subfigure}\hfill
\begin{subfigure}{0.49\textwidth}%
    \includegraphics[width=\linewidth]{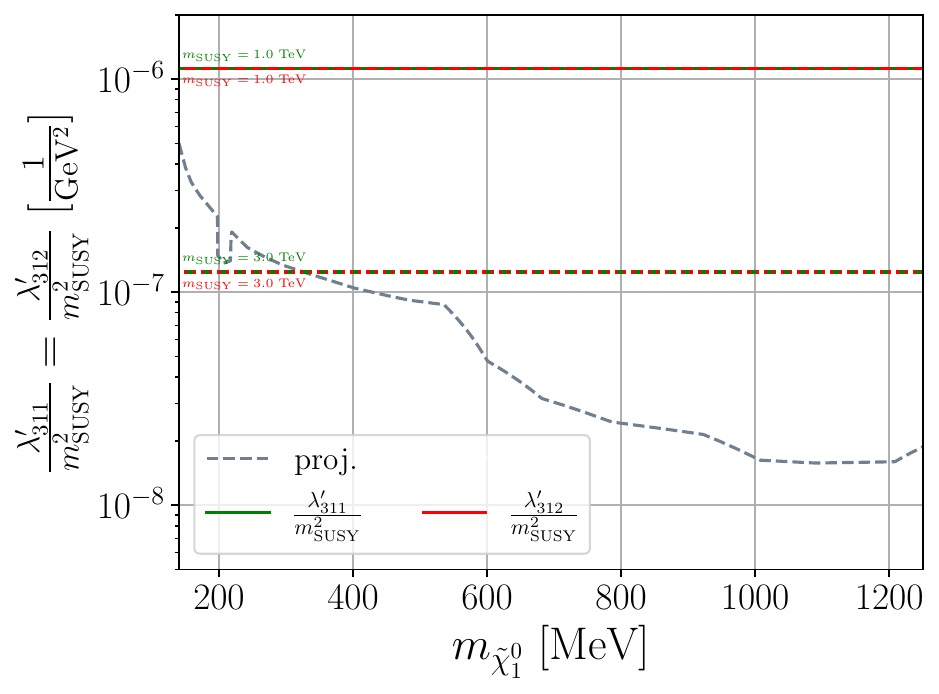}
  \caption{Benchmark $\bm{\tau_2}$ from~\cref{tab:d_BMs}.}
  \label{fig:tau_two_2}
\end{subfigure}
\caption{As in~\cref{fig:kaon_two} but for binos produced from $\tau$ leptons.}
\label{fig:tau_two}
\end{figure}
\begin{figure}[h!]
\begin{subfigure}{0.45\textwidth}
  \includegraphics[width=\linewidth]{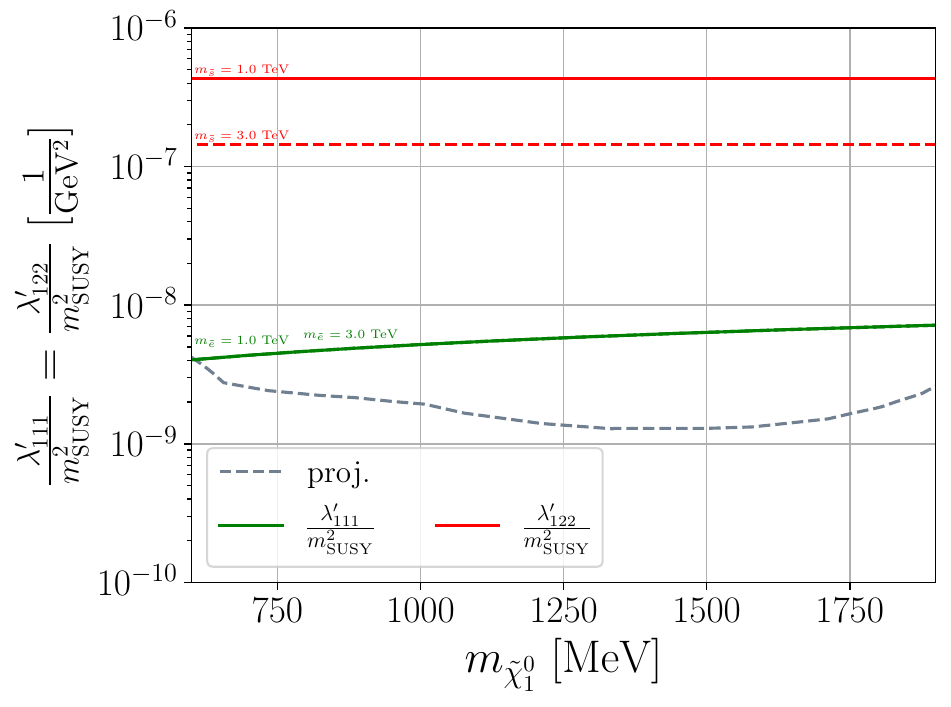}
  \caption{Benchmark $\bm{D_1}$ from~\cref{tab:d_BMs}.}
  \label{fig:D_two_1}
  \end{subfigure}\hfill
\begin{subfigure}{0.45\textwidth}
   \includegraphics[width=\linewidth]{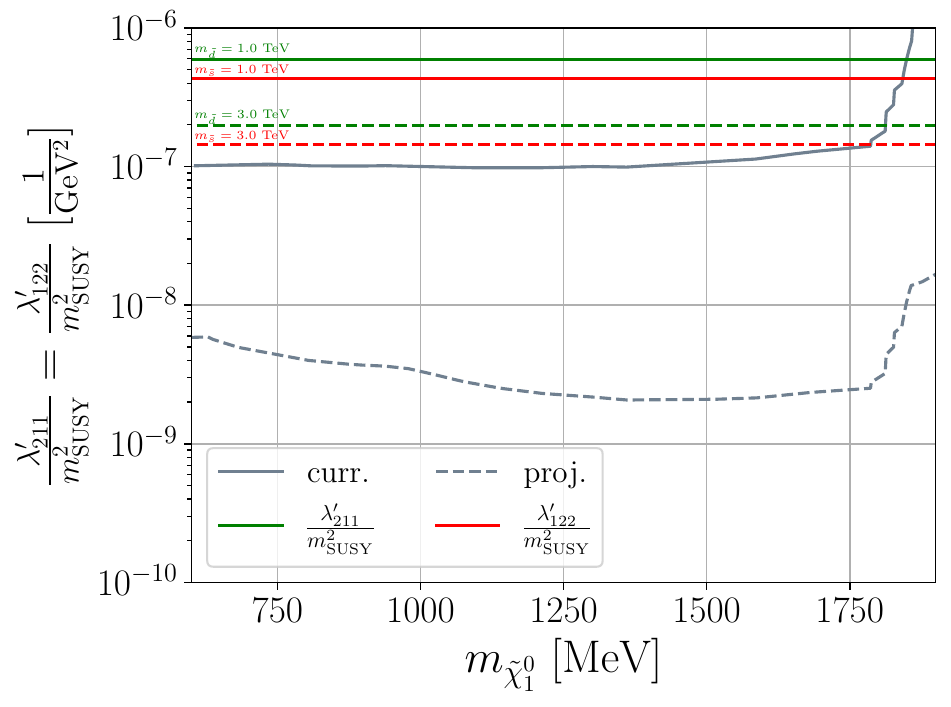}
  \caption{Benchmark $\bm{D_2}$ from~\cref{tab:d_BMs}.}
   \label{fig:D_two_2}
\end{subfigure}\hfill
\\
\begin{subfigure}{0.45\textwidth}%
    \includegraphics[width=\linewidth]{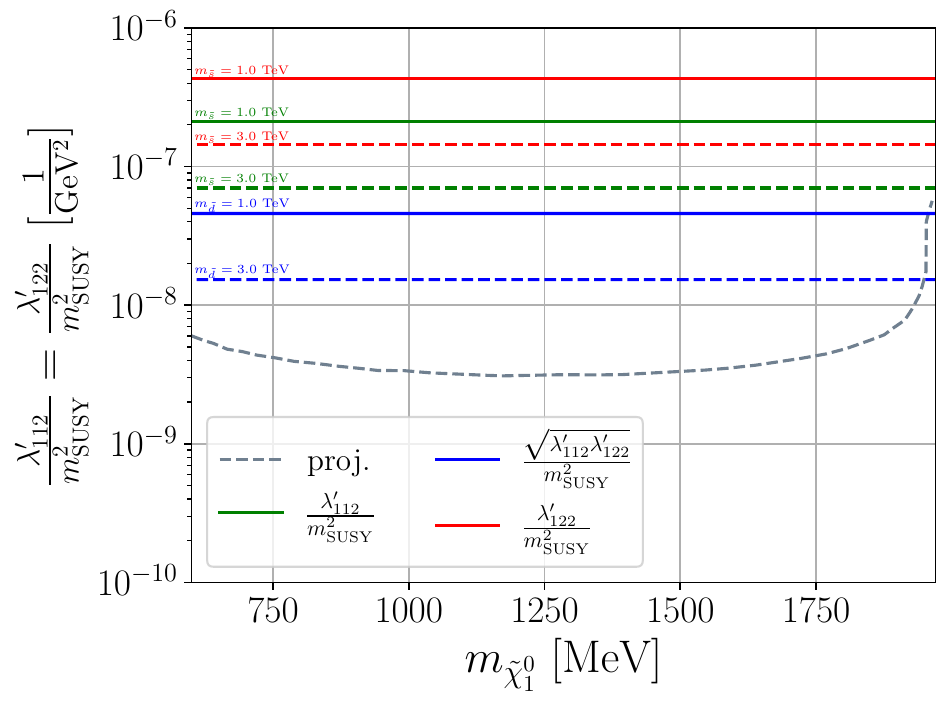}
  \caption{Benchmark $\bm{D_3}$ from~\cref{tab:d_BMs}.}
   \label{fig:D_two_3}
\end{subfigure}\hfill
\begin{subfigure}{0.45\textwidth}
  \includegraphics[width=\linewidth]{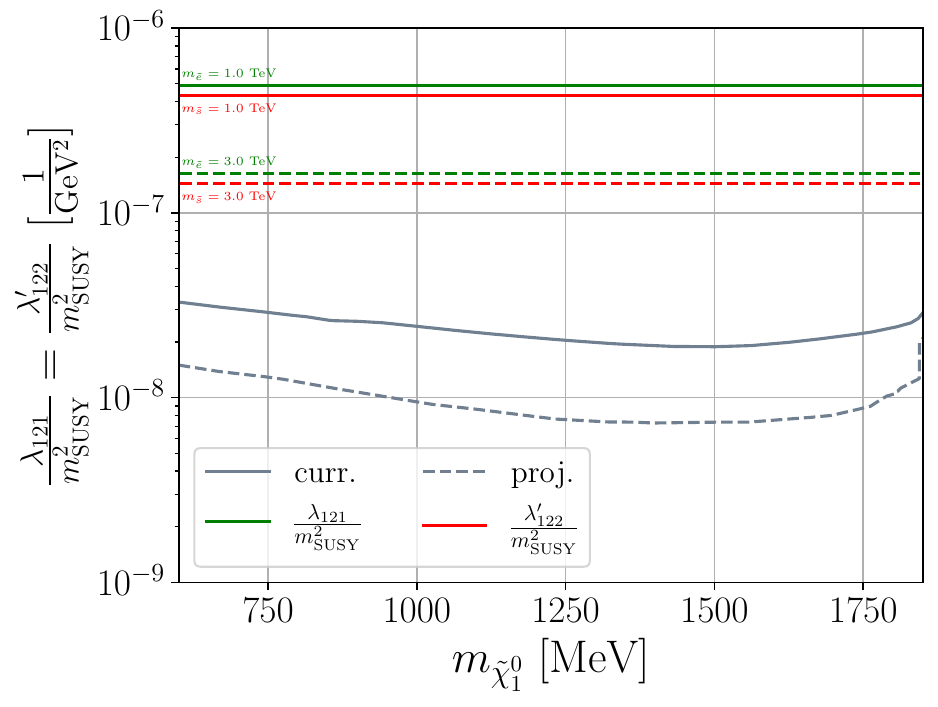}
  \caption{Benchmark $\bm{D_4}$ from~\cref{tab:d_BMs}.}
   \label{fig:D_two_4}
  \end{subfigure}\hfill
  \\
\begin{subfigure}{0.45\textwidth}
   \includegraphics[width=\linewidth]{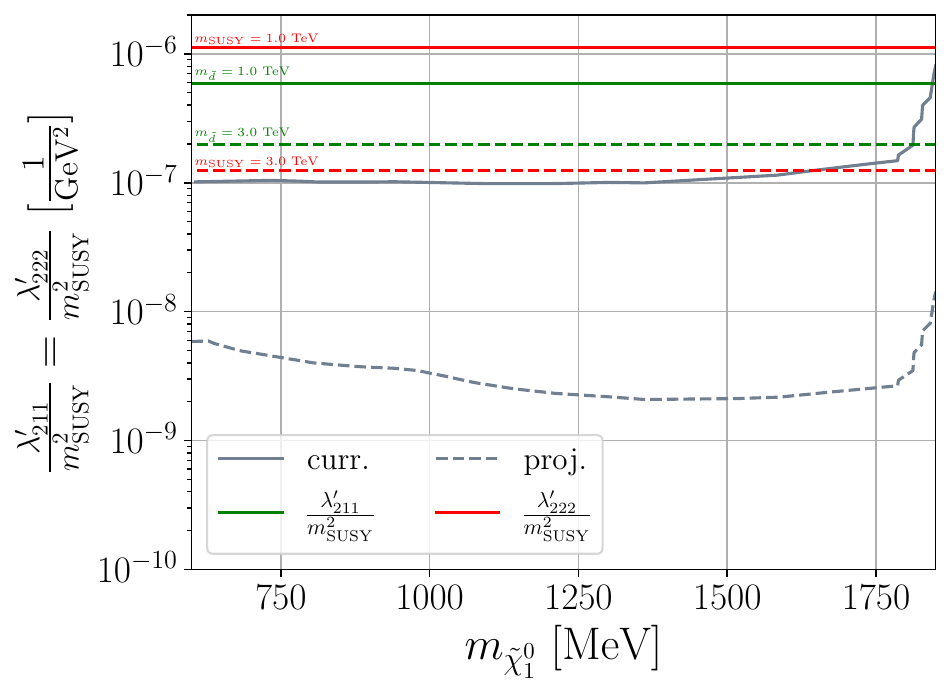}
  \caption{Benchmark $\bm{D_5}$ from~\cref{tab:d_BMs}.}
   \label{fig:D_two_5}
\end{subfigure}\hfill
\begin{subfigure}{0.45\textwidth}%
    \includegraphics[width=\linewidth]{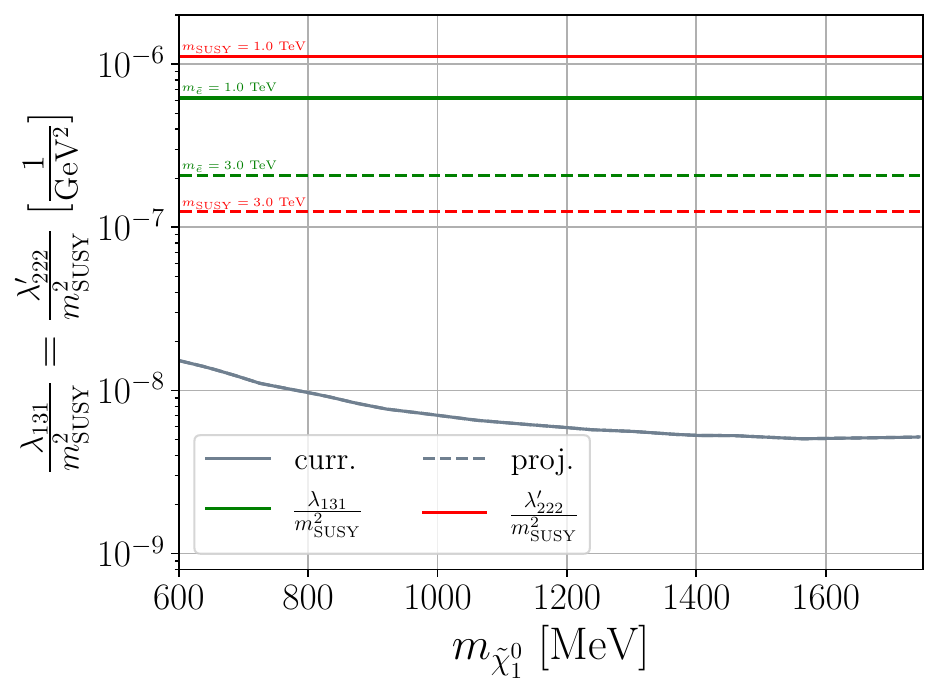}
  \caption{Benchmark $\bm{D_6}$ from~\cref{tab:d_BMs}.}
   \label{fig:D_two_6}
\end{subfigure}
\\
\centering
\begin{subfigure}{0.45\textwidth}
   \includegraphics[width=\linewidth]{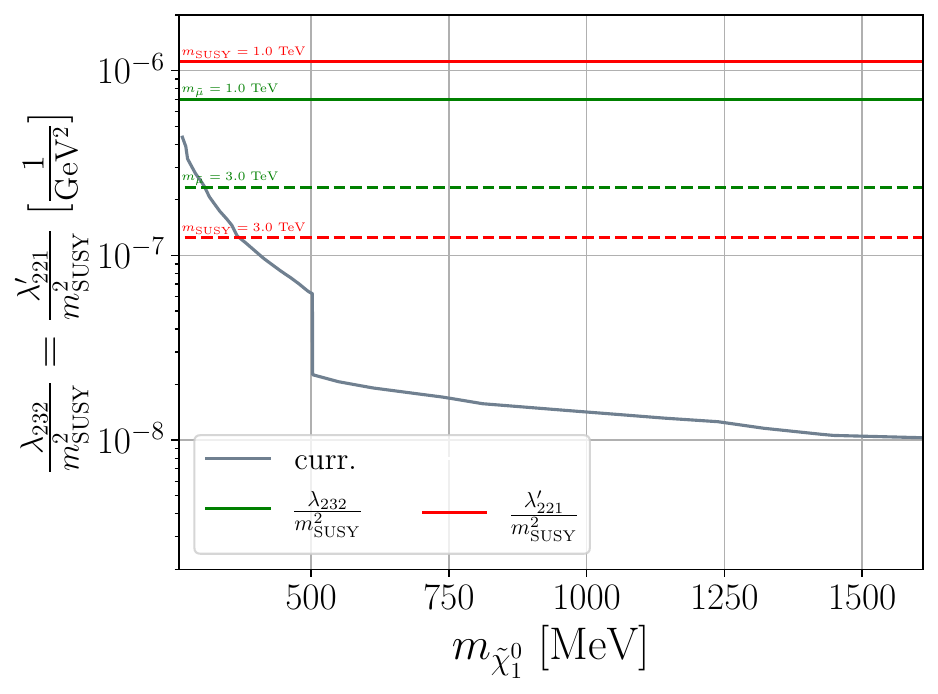}
   \caption{Benchmark $\bm{D_7}$ from~\cref{tab:d_BMs}.}
    \label{fig:D_two_7}
\end{subfigure}
  \caption{As in~\cref{fig:kaon_two} but for binos produced from $D$ and $D_s$ mesons.}
  \label{fig:D_two}
\end{figure}

We summarize the details for the $\tau$ benchmarks we consider,
as well as the $D^\pm$ and $D^\pm_s$ meson ones in~\cref{tab:d_BMs}. We group them 
together in this section since, at the considered experiments, $\tau$ leptons are 
most copiously produced in the decays of the $D_s$ mesons. The corresponding 
sensitivity limits are shown in Fig.~\ref{fig:tau_two} for the $\tau$, and in Fig.~\ref{fig:D_two} for the mesons.
\begin{table}[h]
\centering
\begin{tabular}{|c|c|c|c|}
\hline
\bf{Label} & \bf{Production} & \bf{Decay} & $\bm{m_{\tilde{\chi}^0_1}}$\\
\hline
$\bm{\tau_1}$ & $\lambda'_{311}$ & $\lambda_{212}$ & $\SI{211}{\MeV} - \SI{1637}{\MeV}$\\
$\bm{\tau_2}$ & $\lambda'_{312}$ & $\lambda'_{311}$ & $\SI{140}{\MeV}-\SI{1283}{\MeV}$\\
$\bm{D_1}$ & $\lambda'_{122}$ & $\lambda'_{111}$ & $\SI{600}{\MeV} - \SI{1968}{\MeV}$\\
$\bm{D_2}$ & $\lambda'_{122}$ & $\lambda'_{211}$ & $\SI{600}{\MeV}-\SI{1968}{\MeV}$\\
$\bm{D_3}$ & $\lambda'_{122}$ & $\lambda'_{112}$ & $\SI{600}{\MeV} - \SI{1968}{\MeV}$\\
$\bm{D_4}$ & $\lambda'_{122}$ & $\lambda_{121}$ & $\SI{600}{\MeV}-\SI{1968}{\MeV}$\\
$\bm{D_5}$ & $\lambda'_{222}$ & $\lambda'_{211}$ & $\SI{600}{\MeV} - \SI{1863}{\MeV}$\\
$\bm{D_6}$ & $\lambda'_{222}$ & $\lambda_{131}$ & $\SI{600}{\MeV} - \SI{1863}{\MeV}$\\
$\bm{D_7}$ & $\lambda'_{221}$ & $\lambda_{232}$ & $\SI{260}{\MeV} - \SI{1764}{\MeV}$\\
\hline
\end{tabular}
\caption{As in~\cref{tab:pion_BMs} but for bino production from $\tau$ leptons, and $D$ and $D_s$ mesons.}
\label{tab:d_BMs}
\end{table}

For the $\tau$ lepton scenarios, the exclusion and projected search sensitivity, shown in~\cref{fig:tau_two}, come from \texttt{BEBC}~\cite{WA66:1985mfx, 
Barouki:2022bkt}, \texttt{CHARM}~\cite{CHARM:1985nku, Boiarska:2021yho}, and 
\texttt{ArgoNeuT}~\cite{ArgoNeuT:2021clc}; and \texttt{DUNE}~\cite{Ballett:2019bgd}, 
\texttt{FASER2}~\cite{Kling:2018wct,Beltran:2023nli}, and 
\texttt{MoEDAL-MAPP2}~\cite{DeVries:2020jbs, Beltran:2023nli}, respectively. There is 
no current search targeting the final state of $\bm{\tau_2}$, while the current limit 
on $\bm{\tau_1}$ beats even the projected search sensitivity at 
\texttt{DUNE} and \texttt{FASER}. The sharp drop in sensitivity in~\cref{fig:tau_two_1} 
at $m_{\tilde{\chi}_1^0} \approx \SI{1000}{\MeV}$ occurs because the $\rho$ and bino 
decay mode of the $\tau$ lepton in the RPV model (\textit{cf.}~\cref{tab:single_couplings_BMs}) 
becomes kinematically inaccessible, leading to the reduction in production.

For the $D$ and $D_s$ mesons, current exclusions are provided by searches at 
\texttt{BEBC}~\cite{WA66:1985mfx, Barouki:2022bkt}, \texttt{CHARM}~\cite{CHARM:1985nku}, 
and \texttt{NuTeV}~\cite{NuTeV:1999kej}.
Further, \texttt{DUNE}~\cite{Ballett:2019bgd}, \texttt{FASER2}~\cite{Kling:2018wct}, and \texttt{MoEDAL-MAPP2}~\cite{DeVries:2020jbs,Beltran:2023nli} are projected to improve this reach. For all $D_s$ benchmarks except 
$\bm{D_7}$, we only consider neutralinos with mass, $m_{\tilde{\chi}^0_1} > \SI{600}{\MeV}$.
This is because either the above experiments only constrain the corresponding parameter 
region in the HNL model, or because kaons and pions dominate the HNL production for 
lower masses. One exception is \texttt{BEBC}, where nearly all produced pions and kaons 
are absorbed by a high-density target before they can decay, and HNL production is 
dominated by $D$ meson decays. Since benchmark $\bm{D_7}$ involves production from 
$D$-mesons, we can probe lower bino masses in our scenario.
The kink in the corresponding sensitivity limit in~\cref{fig:D_two_7} at $m_{\tilde{\chi}_1^0} \approx \SI{500}{\MeV}$ occurs because \texttt{CHARM} takes over from \texttt{BEBC}.
We note that the final states of benchmarks $\bm{D_1}$ and $\bm{D_3}$ are not covered by the existing searches but will be covered by the upcoming experiments.

Once again, we see -- for all the plots in this section -- that reinterpreting existing 
and projected limits on HNL models in terms of our RPV scenarios gives bounds on the 
parameter space that improve upon existing limits by orders of magnitude.

\subsubsection{$B$ and $B_c$ scenarios} 
Here, the bino is produced in $B (B^\pm/\overset{\scriptscriptstyle(-)}{B^0})$ or $B^\pm_c$ 
decays, \textit{cf.}~\cref{tab:b_BMs}. While HNL production in these 
modes only becomes dominant above the $D$-meson thresholds, the projected search 
sensitivity for \texttt{FASER(2)} provided in Ref.~\cite{Kling:2018wct} shows the results separately for the HNLs from $B$-meson decays and from $D$-meson and kaon 
decays, enabling us to choose benchmarks with masses lower than the $D$ thresholds. Up 
to about $m_N \approx \SI{2700}{\MeV}$, the production is dominated by $B$ decays and we choose 
the first three benchmarks accordingly. Beyond this, $B_c$ decays also become 
significant, and for $m_N \gsim \SI{3500}{\MeV}$, they become the dominant modes; 
the last two benchmarks focus on this.

\begin{table}[h]
\centering
\begin{tabular}{|c|c|c|c|}
\hline
\bf{Label} & \bf{Production} & \bf{Decay} & $\bm{m_{\tilde{\chi}^0_1}}$\\
\hline
$\bm{B_1}$ & $\lambda'_{113}$ & $\lambda'_{122}$ & $\SI{548}{\MeV} - \SI{2700}{\MeV}$\\
$\bm{B_2}$ & $\lambda'_{113}$ & $\lambda_{131}$ & $\SI{160}{\MeV}-\SI{2700}{\MeV}$\\
$\bm{B_3}$ & $\lambda'_{213}$ & $\lambda'_{211}$ & $\SI{160}{\MeV} - \SI{2700}{\MeV}$\\
$\bm{B_4}$ & $\lambda'_{123}$ & $\lambda'_{311}$ & $\SI{3500}{\MeV}-\SI{6275}{\MeV}$\\
$\bm{B_5}$ & $\lambda'_{123}$ & $\lambda_{131}$ & $\SI{3500}{\MeV} - \SI{6275}{\MeV}$\\
\hline
\end{tabular}
\caption{As in~\cref{tab:pion_BMs} but for bino production from $B$ and $B_c$ mesons.}
\label{tab:b_BMs}
\end{table}

\begin{figure}[h]
\begin{subfigure}{0.48\textwidth}
  \includegraphics[width=\linewidth]{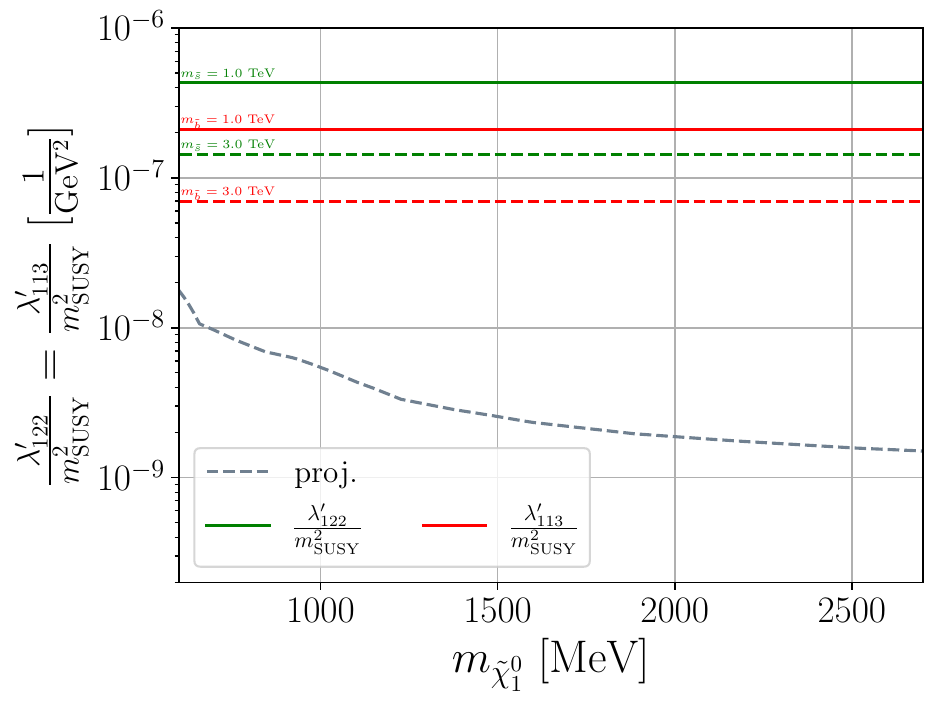}
  \caption{Benchmark $\bm{B_1}$ from~\cref{tab:b_BMs}.}
  \end{subfigure}\hfill
\begin{subfigure}{0.48\textwidth}%
    \includegraphics[width=\linewidth]{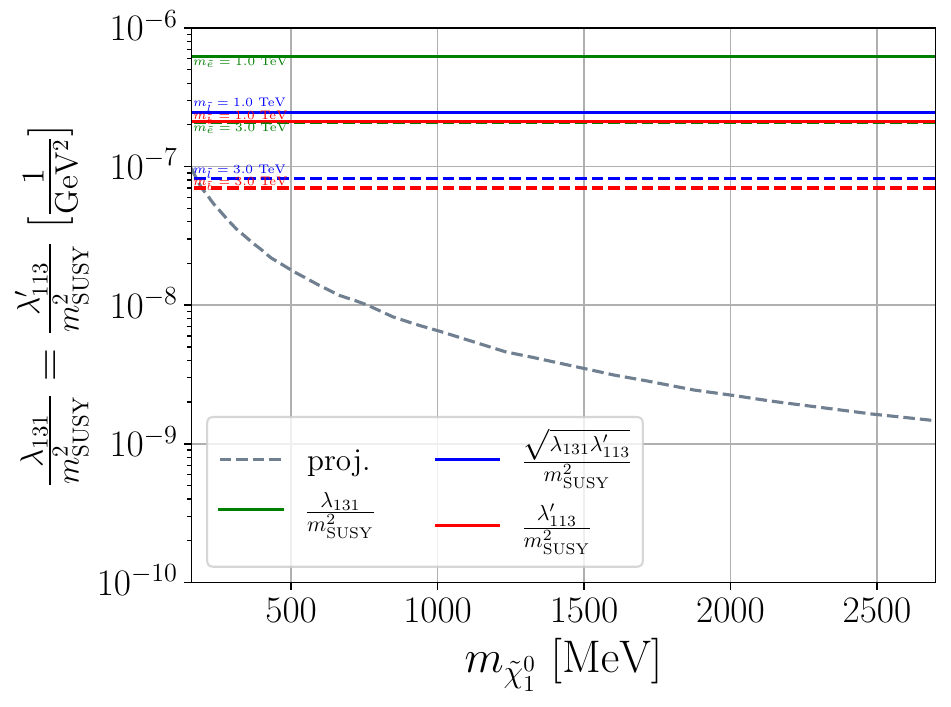}
  \caption{Benchmark $\bm{B_2}$ from~\cref{tab:b_BMs}.}
\end{subfigure}
\\
\begin{subfigure}{0.48\textwidth}
  \includegraphics[width=\linewidth]{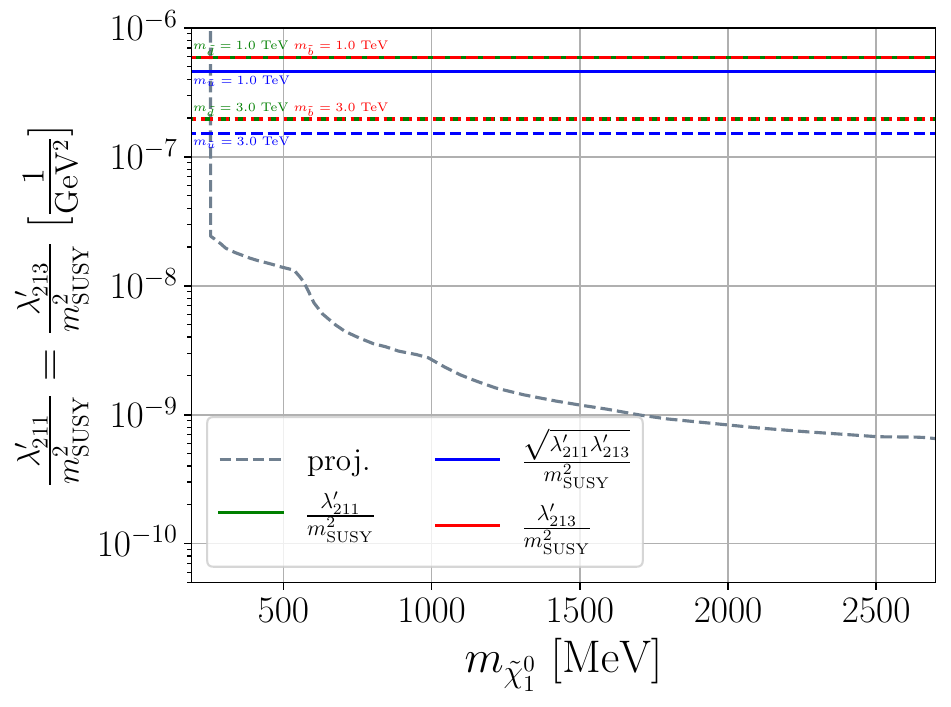}
  \caption{Benchmark $\bm{B_3}$ from~\cref{tab:b_BMs}.}
  \end{subfigure}\hfill
\begin{subfigure}{0.48\textwidth}%
    \includegraphics[width=\linewidth]{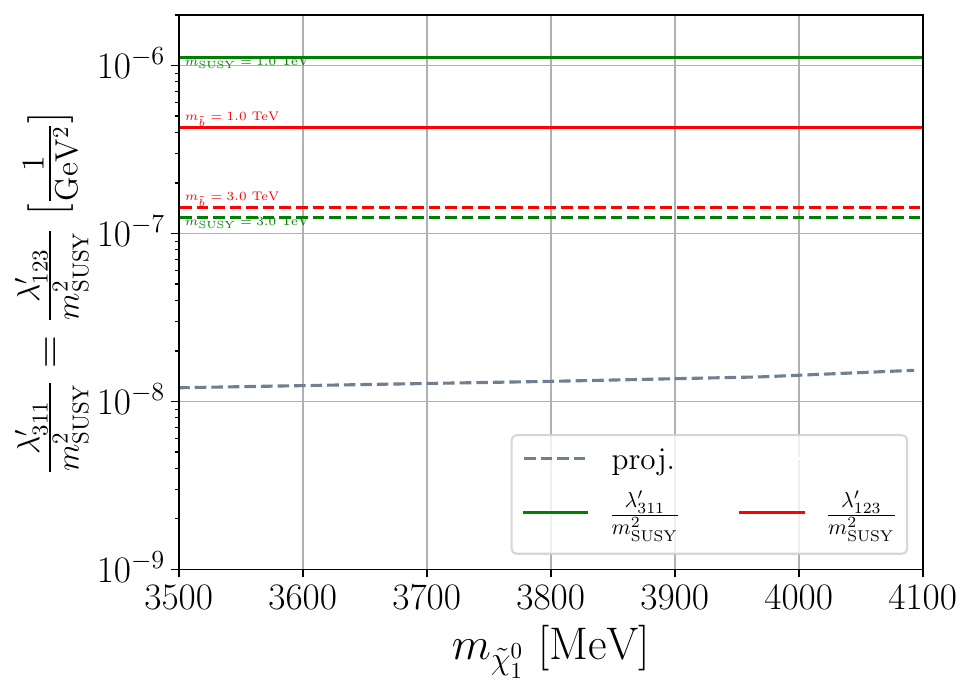}
  \caption{Benchmark $\bm{B_4}$ from~\cref{tab:b_BMs}.}
\end{subfigure}
\\
\centering
\begin{subfigure}{0.48\textwidth}
  \includegraphics[width=\linewidth]{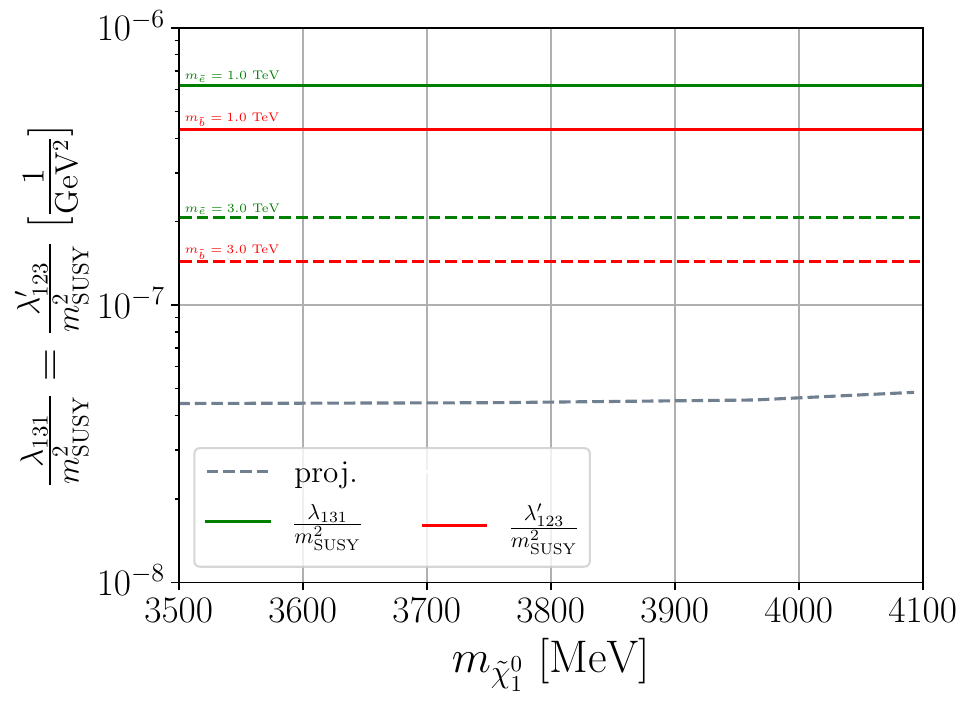}
  \caption{Benchmark $\bm{B_5}$ from~\cref{tab:b_BMs}.}
  \end{subfigure}
\caption{As in~\cref{fig:kaon_two} but for binos produced from $B$ and $B_c$ mesons.}
\label{fig:B_two}
\end{figure}

The corresponding sensitivity limits are presented in Fig.~\ref{fig:B_two}.
There are no existing constraints; however, projections from \texttt{FASER2}~\cite{Kling:2018wct,Beltran:2023nli} and \texttt{MoEDAL-MAPP2}~\cite{DeVries:2020jbs,Beltran:2023nli} show that we should be able to probe the RPV parameter space up to 2-3 orders of magnitude beyond what is ruled out by current limits.
\section{Conclusions}\label{sec:conclusions}

In this work, we have considered a GeV-scale or lighter long-lived lightest neutralino, 
which is necessarily bino-like, in the minimal R-parity-violating (RPV) supersymmetric model.
We have focused on lepton-number-violating operators in the RPV superpotential: $LH_u$, 
$LL\bar{E}$, and $LQ\bar{D}$. Such light neutralinos are still allowed by all 
experimental and observational constraints, as long as they decay, for instance via RPV couplings, 
so as to avoid overclosing the Universe. Since these RPV couplings are bounded to be small, such 
light binos, which we assume to be the lightest supersymmetric particle in our theory, are expected 
to have a relatively long lifetime. Via the considered couplings, the binos can decay leptonically 
or semi-leptonically. For their production, we have focused on rare decays of 
mesons and charged leptons which are copiously produced at various facilities such as beam-dump and 
collider experiments. Once produced, these light binos can lead to exotic signatures such as 
displaced vertices (DVs) or missing energy.
We have used searches for these signatures to constrain the RPV couplings associated with a light bino.

These various strategies are experimentally widely utilized to constrain heavy neutral leptons 
(HNLs) which may decay to almost the same final-state particles as the lightest neutralino in the 
RPV models. We have thus used the existing HNL searches to set new strict bounds on the relevant 
RPV couplings.
Furthermore we have translated the prjected sensitivity to the HNL parameters at certain future experiments into the corresponding search sensitivity for the light neutralinos.

We have studied comprehensively the past experiments \texttt{PIENU}, 
\texttt{NA62}, \texttt{T2K}, and \texttt{BaBar}, as well as the approved experiments \texttt{FASER}, \texttt{MoEDAL-MAPP1}, \texttt{PIONEER}, and \texttt{DUNE}.
We did not consider future experiments that are not yet approved such as \texttt{MATHUSLA} and \texttt{ANUBIS}, with two exceptions, namely \texttt{FASER2} and \texttt{MoEDAL-MAPP2} since they would be the follow-up programs of the two currently 
running experiments \texttt{FASER} and \texttt{MoEDAL-MAPP1}.

Given the various types and flavor-indices of the RPV operators that can be switched on, we have 
investigated separately different theoretical benchmark scenarios, which can be bounded by distinct 
experiments and strategies. For the selected representative benchmark scenarios, we have performed 
numerical computation and presented the final exclusion bounds. Further, we have compared these 
recast bounds with the existing limits on the RPV couplings which mainly stem from low-energy 
processes of meson and lepton decays. In general, we find that in most cases the exclusion limits 
obtained from recasting past HNL searches surpass the existing bounds on the RPV 
couplings by orders of magnitude, and the expected limits at the considered ongoing and future 
experiments can be even stronger.

Simple and analytic reinterpretation methods are becoming an important research tool.
This is because most published experimental reports present results only for a limited number of simple models, and a recast with full simulation is often complicated and time-consuming.
Some existing works such as Refs.~\cite{Fernandez-Martinez:2023phj, Barouki:2022bkt, Beltran:2023nli} have shown the power of simple and quick reinterpretation of searches for long-lived particles by considering heavy neutral leptons in various models and axion-like particles as examples.
Our work exemplifies again the strengths and convenience of such reinterpretation methods, by recasting the bounds on the HNLs in the minimal scenario, into those on the lightest neutralinos in the RPV supersymmetric models, and hence motivates the development of further studies with these reinterpretation methods.

\bigskip
\textbf{Addendum:}\\ The authors note that Julian Günther et al.~are currently working on a dedicated simulation for the same benchmark $\bm{K_1}$ at \texttt{DUNE} in a soon-to-be-released paper.
\bigskip
\section*{Acknowledgments}
\bigskip
We thank Subir Sarkar and Giacomo Marocco for initial stimulating discussions on their
recasting of the \texttt{BEBC} results. We thank Rhorry Gauld for contributions in 
the early stage of this project, and thank Julian Günther and Martin Hirsch for 
useful discussions. HKD, DK, SN, and MS acknowledge partial ﬁnancial support by the 
Deutsche Forschungsgemeinschaft (DFG, German Research Foundation) through the funds 
provided to the Sino-German Collaborative Research Center TRR110 “Symmetries and the 
Emergence of Structure in QCD” (DFG Project ID 196253076 - TRR 110).

\bigskip
\bigskip
\appendix
\section{Explicit neutralino production/decay widths with $LL\bar{E}$ operators}\label{sec:appendixA}
In this appendix, we will give the explicit general formulae needed for both the neutralino production and decay via $LL\bar E$ couplings at tree level. 
In the framework of the RPV-MSSM, the relevant processes will always involve four external two-component fermions and one intermediate scalar. The external fermions may carry momenta $p_i$ and have masses $m_i$, where $i$ is a generic label with $i = 0$ (incoming) and $i=1,2,3$ (outgoing). Following Ref.~\cite{Dreiner:2008tw}, we express the total decay width as 
\begin{equation}
    \label{eq:generic_LLE_width}
    \varGamma_{LL\bar E} (0;1,2,3)[\alpha,\beta,\gamma] = \frac{m_0}{2^8\pi^3} \int_{z_{3}^\text{min}}^{z_{3}^\text{max}} \d z_{3} \int_{z_1^\text{min}}^{z_1^\text{max}} \d z_1 \:\overline{|\mathcal{M}|^2} \, , \\
\end{equation}
where the spin-averaged matrix element takes the form
\begin{align}
    \label{eq:LLE_ME}
    \overline{|\mathcal{M}|^2} &= \frac{m_0^4}{2} \bigg[ |\alpha|^2 \mathcal{Z}_1 + |\beta|^2 \mathcal{Z}_2 + |\gamma|^2 \mathcal{Z}_3 \nonumber \\
    & \qquad \qquad  - \Re\{\alpha \beta^*\} \Big(+\mathcal{Z}_1 + \mathcal{Z}_2 - \mathcal{Z}_3 \Big) \nonumber\\
    & \qquad \qquad  - \Re\{\beta \gamma^*\} \Big(-\mathcal{Z}_1 + \mathcal{Z}_2 + \mathcal{Z}_3 \Big) \nonumber \\
    & \qquad \qquad  - \Re\{\alpha \gamma^*\} \Big(+\mathcal{Z}_1 - \mathcal{Z}_2 + \mathcal{Z}_3 \Big)\bigg]\, ,
\end{align}
with
\begin{equation}
    \mathcal{Z}_i \equiv z_i \left(1-z_i+2\xi_i^2-\sum_{j=1}^3 \xi_j^2\, \right).
\end{equation}
The kinematic variables $z_{i}$ are defined as
\begin{align}
    z_{i} &\equiv 2p_0\cdot p_{i} / m_0^2 = 2E_{i}/m_0  \, ,
\end{align}
and fulfill the relation $\sum_{i=1}^3 z_i = 2$. Furthermore we introduce the mass ratios
\begin{equation}
    \xi_{i} \equiv \frac{m_{i}}{m_0} \quad (i \neq 0)\, .
\end{equation}
The (in general) complex valued coefficients $\alpha, \beta, \gamma$ follow from the Feynman rules relevant for the respective process, and can be simplified with the assumption of degenerate sfermion masses, \textit{cf}.~Sec.~\ref{sec:model}.
Their explicit expressions are given in Eq.~\eqref{eq:coeff_LLE}.
The integration limits in Eq.~(\ref{eq:generic_LLE_width}) can be obtained from the minimal and maximal values of the invariant masses of the $``1-2$'' and $``2-3$'' systems, \textit{cf}.~Ref.~\cite{ParticleDataGroup:2022pth}:
\begin{align}
    (m_{12}^2)_\text{max} &= (m_0-m_{3})^2\, , \label{eq:inv_masses1}\\
    (m_{12}^2)_\text{min} &= (m_1+m_{2})^2 \, , \\
    (m_{23}^2)_\text{max} &= (E_{2}^*+E_{3}^*)^2 - \left(\sqrt{{E_{2}^*}^2-m_{2}^2}-\sqrt{{E_{3}^*}^2-m_{3}^2}\right)^2 \, ,\\
    (m_{23}^2)_\text{min} &= (E_{2}^*+E_{3}^*)^2 - \left(\sqrt{{E_{2}^*}^2-m_{2}^2}+\sqrt{{E_{3}^*}^2-m_{3}^2}\right)^2 \, ,
\end{align}
where the energies $E_{2,3}^*$ in the $``1-2$'' rest frame are given in terms of $m_{12}$ by,
\begin{align}
    E_{2}^* &= (m_{12}^2-m_1^2+m_{2}^2)/2m_{12} \, ,\\
    E_{3}^* &= (m_0^2 - m_{12}^2-m_{3}^2)/2m_{12} \label{eq:E_star_minus}\, .
\end{align}
From energy-momentum conservation one can deduce that
\begin{equation}
    z_{3} = \frac{m_0^2+m_{3}^2 - m_{12}^2}{m_0^2} \qquad \text{and} \qquad z_{1} = \frac{m_0^2 +m_1^2 - m_{23}^2}{m_0^2}\, ,
\end{equation}
which then finally yield the integration limits:
\begin{align}
    z_{3}^\text{max} &= 1 + \xi_{3}^2 - \frac{(m_{12}^2)_\text{min}}{m_0^2} \, ,\\
    z_{3}^\text{min} &= 1 + \xi_{3}^2 - \frac{(m_{12}^2)_\text{max}}{m_0^2} = 2\xi_{3} \, ,\\
    z_1^\text{max} &= 1 + \xi_{1}^2 - \frac{(m_{23}^2)_\text{min}}{\mchi^2} \, ,\\
    z_1^\text{min} &= 1 + \xi_{1}^2 - \frac{(m_{23}^2)_\text{max}}{\mchi^2}\, .
\end{align}

\bibliographystyle{JHEP}
\bibliography{refs}
\end{document}